\documentclass[english, aps, pra, preprint, onecolumn]{revtex4-1}
\usepackage[T1]{fontenc}
\usepackage{lmodern}
\usepackage{verbatim}
\usepackage{bm}
\usepackage{amsmath}
\usepackage{amssymb}

\usepackage{graphicx}
\usepackage{xcolor}
\usepackage{color}

\usepackage{braket}
\usepackage{subcaption}

\usepackage{hyperref}

\begin{document}

\preprint{Three level notes}
\title{Entanglement of a three-level atom interacting with two-modes field in a cavity}
\author{A. del R\'{i}o-Lima}
\affiliation{Instituto de F\'{i}sica, Universidad Nacional Aut\'{o}noma de M\'{e}xico, 0100 CDMX,  M\'{e}xico}
\author{F. J. Poveda-Cuevas}
\affiliation{C\'{a}tedras Conahcyt - Instituto de F\'{i}sica, Universidad Nacional Aut\'{o}noma de M\'{e}xico, 0100 CDMX,  M\'{e}xico}
\author{O. Casta\~{n}os}
\affiliation{Instituto de Ciencias Nuclares, Universidad Nacional Aut\'{o}noma de M\'{e}xico, 04510 CDMX, México}
\begin{abstract}
The dynamics of the interaction between an atom of three levels interacting with a quantized field of two modes in a cavity is studied within the rotating wave approximation, by taking into account experimental values of the accessible hyperfine levels of alkaline atoms. An equal detuning is considered to determine the matter-field entanglement, the statistical properties of the photons, and the occupation probabilities of the atom.  For a large detuning or weak dipolar strength  appear the Raman condition, that is, the suppression of one of his atomic transitions. Analytic expression for the time evolution operator allows to have also explicit closed expressions for the field and matter observables. 
\end{abstract}
\maketitle

\section{Introduction}

Entanglement phenomena was recognized since the foundations of quantum mechanics \cite{Einstein1935, Schroedinger1935}, which has lead to experimental confirmable deviations of statistical correlations of quantum states from classical concepts of reality and locality \cite{Bell2004, Clauser1969, Aspect1982}.  The entanglement is now recognized as a useful resource to yield quantum teleportation of states, to do tasks of quantum cryptography, and to control quantum systems. For this reason, there are an increasing interest in the entanglement theory and in establishing experimental devices to generate entanglement \cite{Guehne2009, Barnett2017, Barnett1989, Barnett2009}. 

One of this devices can be generated by Hamiltonian models of many levels atom interacting with photon modes in a cavity. Because photons and atoms confined in cavities have a behavior showing some of the fundamental principles of the quantum physics and make possible the development of new measurement devices \cite{Haroche2013, Pike1986}. 

In the eighties a large quantity of works were published related with theoretical and experimental studies of matter-light interactions. These works consider one atom with a few active energy levels interacting with one or more electromagnetic modes in a resonant cavity \cite{Yoo1985, Pike1986, Kaluzny1983, Rempe1987, Raizen1989}.

The paradigm of the Hamiltonian models is associated to the Jaynes-Cummings model \cite{Jaynes1963} in which one two-level atom interacts with one mode of electromagnetic radiation in a cavity. This model can be solved analytically in the rotating wave approximation (RWA) and its dynamical behavior has lead to the prediction of the revivals and collapses of the Rabi oscillations of the atomic occupation probability \cite{Eberly1980, Narozhny1981}. This phenomena were later demonstrated experimentally by means of a one-atom maser \cite{Meschede1985}. Multiphoton process has also been investigated, in particular the two-photon Jaynes-Cummings model which is also exactly solved and exhibits also the collapse and revival phenomena \cite{Gerry1988, He1989}.

Hamiltonian models with one atom with many levels interacting with one, two, or more modes of the electromagnetic field in a cavity have also been studied \cite{Yoo1985, Stenholm1973, Jr1986, Li1985, Zhu1989}, and finding collapse and revivals in the atomic inversion observable and other results analogous to those appearing in the standard  Jaynes-Cummings model.

More recently these models of matter field Hamiltonians have been used to shown the dynamic entanglement phenomena between the matter and the field.  The JCM is considered to study the fluctuations in the quantum evolution of pure states by calculating variances, the von Neumann entropy and the Shannon entropy \cite{Phoenix1988}, although there is not an explicit mention of the entanglement phenomena between the electromagnetic field and the atom, neither that the entanglement present can be used as a resource in quantum information theory. The analysis is extended by considering the evolution of initial pure and mixed states for the light and the matter sectors \cite{Boukobza2005}.

The dynamics of a three level atom in the lambda configuration interacting with two modes of the electromagnetic field with the upper level removed when there is a large detuning (far from resonance and it is called adiabatic elimination) constitutes an effective model of the Raman scattering involving the pump (mode photon one) and the Stokes (photon mode 2) modes \cite{Gerry1990}.  We should note here that a one-mode Raman coupled model involving two degenerate atomic states connected by a single-mode two-photon coupling has been discussed in relation with the collapses and revivals phenomena \cite{Knight1986, Phoenix1990, Gou1989}. The study of the same model, the three level atom in the lambda configuration interacting with two modes, but without the removal of one level has been also investigated in \cite{Bogolubov1986, Bogolubov1987}, where non-classical properties of light were shown. 

Notice that, for a two-level atom interacting with two modes of fields in a cavity, which are called two-photon processes, the occupation probabilities as function of time show the collapses and revivals phenomena, which are due to the granular nature of the field and the revivals are regular and compact. In these systems, it is assumed that the two levels have the same parity and the initial states describing  the modes of the field in the cavity were uncorrelated or that the number of photons in one mode is equal to the number of photons in the second mode \cite{Gou1989}. It was also shown that the dynamics is different to the case of single-mode Jaynes-Cummings models discussed by Eberly et al., where the revivals are regular but overlapping in an initial  coherent state and irregular in an initial chaotic field \cite{Eberly1980, Narozhny1981}. For initial states with different number of photons, called pair-coherent states or correlated SU$(1,1)$ coherent states, which has  exceptional features such as sub-Poissonian statistics, squeezing, and violations of Cauchy-Schwarz inequalities, the study of the collapses and revivals of the Rabi oscillations in a field was realized in \cite{Agarwal1988, Joshi1990, Gerry1992}. It is important to mention that the collapse and revival phenomena predicted by the Jaynes-Cummings model have been observed experimentally in a one-atom maser (a single two-level atom interacting with a single-mode electromagnetic field) \cite{Rempe1987, Meschede1985}. 

The quantum phase transitions constitutes another area of research in quantum optics, where the Jaynes-Cummings model and their extensions or generalizations has been frequently used \cite{NahmadAchar2015}. In particular the quantum phase transitions associated to Hamiltonian models describing the interaction of three level atoms interacting dipolarly with two electromagnetic modes in a cavity have been determined recently, where the entanglement properties between the matter and field sectors have played a fundamental role in the calculation of the corresponding quantum phase diagrams \cite{Cordero2021, Castanos2022}. The ground state  of a single atom in the $\Lambda$ configuration exhibits some regions in the dipolar interaction strengths where Wigner functions take negative values. Then in those regions the system shows a non-classical behavior, which are of practical interest in quantum computing and quantum metrology \cite{LopezPena2021}.

\begin{figure}
\begin{centering}
\includegraphics[scale=0.6]{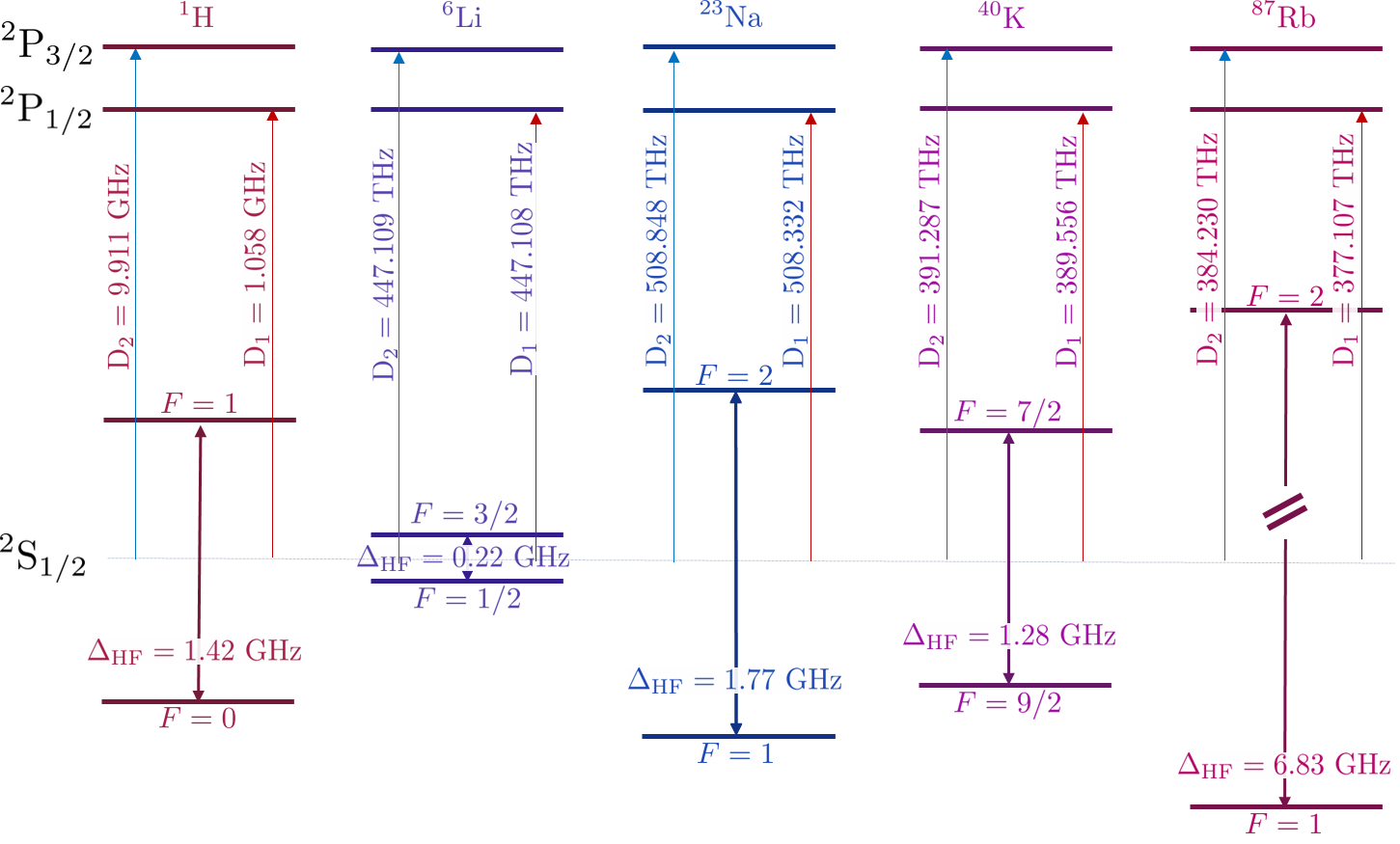}
\end{centering}
\caption{\label{fig:H-structure} Hydrogen and alkaline atoms are characterized for its simplified hyperfine electronic structure, as it is shown.  The $D_1$ and $D_2$ transition lines are also indicated, together with its corresponding spectroscopic notation.}
\end{figure}

Here we are going to study the entanglement properties of general wave packets evolving within the framework of a Hamiltonian model describing the interaction of a three level atom in the $\Lambda$ configuration interacting dipolarly with two electromagnetic modes in a cavity. By considering the experimental information available in the hyperfine energy levels of $^6$Li and $^{87}$Rb. Experimental determination of the hyperfine structure can be found in \cite{Arimondo1977}. This information will allow us determine the energy levels $\hbar\omega_0$, $\hbar\omega_1$ and $\hbar\omega_3$ of the atoms, the frequencies of the quantized fields $\Omega_1$ and $\Omega_2$ in the cavity and mainly the effective dipolar couplings $\mu_{13}$ and $\mu_{23}$ between the atom and the electromagnetic modes. Therefore the evolution of the field and matter observables are computed using the experimental data of alkaline atoms, which commonly are used in ultracold atomic experiments for precision measurements and in many quantum simulations. Alkaline atoms have a  similar electronic structure to the hydrogen atom,  as it is exhibited in Fig.~\ref{fig:H-structure}. The ground state level $^2S_{1/2}$ is split into two hyperfine levels for these types of atoms when the nuclear spin $I$ is taken into account. A fine transition can be made via commercial lasers to excited states $^2P_{1/2}$ and $^2P_{3/2}$. Transition $^2S_{1/2} \rightarrow ^2P_{1/2}$ is known as $D_1$. Transition $^2S_{1/2} \rightarrow ^2P_{3/2}$ is known as $D_2$. Both transitions satisfy the following selection rules $\Delta J=\pm1$ and $\Delta F=0,\pm1$. The three-level system is constructed by choosing a fine transition as reference ($D_1$ or $D_2$) and making the transitions between the hyperfine levels.

This system is used to determine the associated collapses and revivals, the photon statistics for the modes of the field, and their conduct in phase space, for different initial states describing the two-mode electromagnetic field of the cavity.  Furthermore, the dynamic entanglement of pure states by means of the linear or von Neumann entropies is calculated, when one is considering the quantum correlations between the atom and the two-modes of the electromagnetic field.  Additionally we establish the dynamic conditions to have a two-level system with an effective Raman coupling, which depends on the strengths of the dipolar couplings between the atomic levels.

The outline of the contribution is the following: In the second section, we consider a Hamiltonian system constituted by one atom, in the $\Lambda$ atomic configuration, interacting dipolarly with two-modes of electromagnetic radiation in a cavity, within the rotating wave approximation.  The dressed states of the system are calculated in the third section by means of the diagonalization of a three dimensional Hermitean matrix. In the fourth section we construct the general form of the evolution operator of the Hamiltonian system and determine analytically the wave packet at an arbitrary time $t$.  In the fifth section, the entanglement and coherences properties of this wave packet at time $t$ are calculated, together with other expectation values of matter and field observables. Final remarks are given in the conclusions. 

\section{Hamiltonian Model for the atomic $\Lambda$ configuration}

\begin{figure}
\begin{centering}
\includegraphics[scale=0.4]{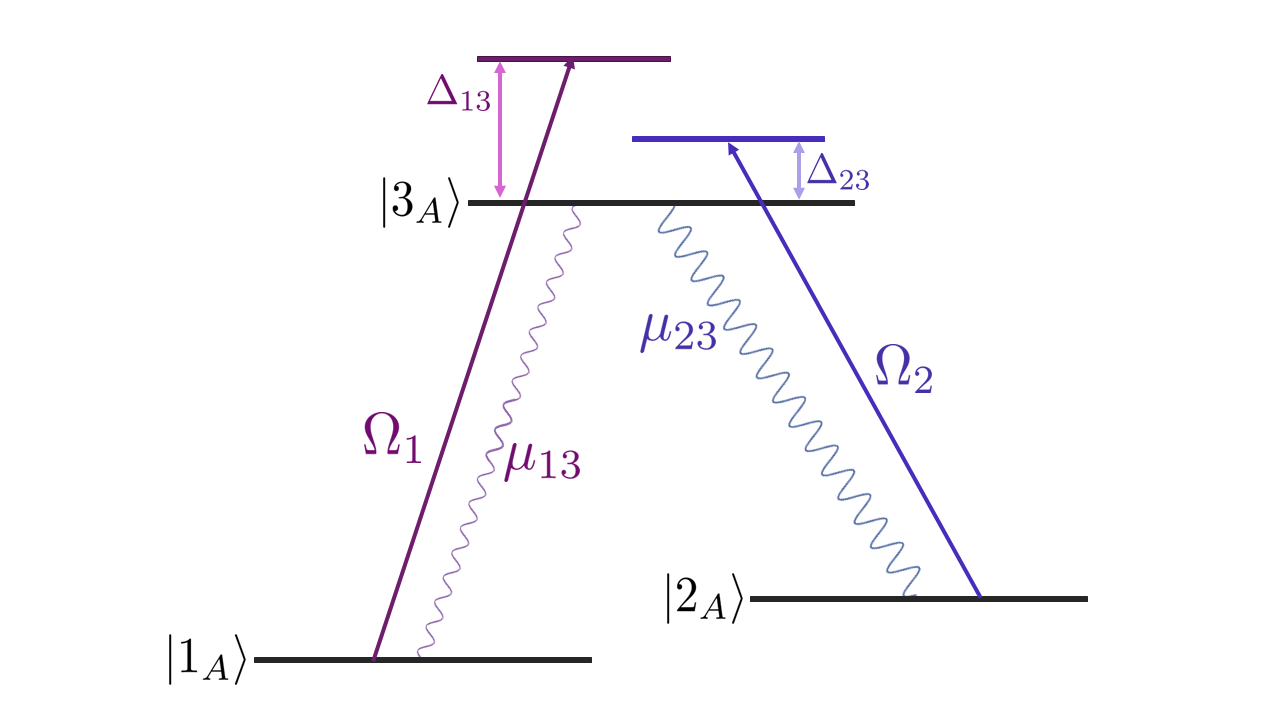}
\end{centering}
\caption{\label{fig:3:level_model} Scheme of an atom of three level with states $\{ \ket{1_A},\ket{2_A},\ket{3_A}\}$. The model considers dipolar transitions, thus transition between states $\ket{1_A}$ and $\ket{2_A}$ is forbidden. The only allowed transitions are $\ket{1_A}\rightarrow \ket{3_A}$ and $\ket{2_A}\rightarrow \ket{3_A}$, with photons of frequencies $\Omega_1$ and $\Omega_2$, respectively.   }
\end{figure}

The Hamiltonian model of an effective
three-level atom in the $\Lambda$ configuration interacting dipolarly with a two-mode electromagnetic field in a cavity (Figure \ref{fig:3:level_model}), in the rotating wave approximation, can be written as follows
\begin{eqnarray}
\hat{H}  &=&  \sum^3_{k=1}  E_{k}\, \hat{A}_{kk}+ \sum^2_{s=1}\hbar\, \Omega_{s} \, \hat{n}_{s} -  \hbar\, \mu_{13}\, \left(\hat{a}_{1}^{\dagger}\hat{A}_{13}+\hat{a}_{1}\hat{A}_{31}\right)-\hbar\, \mu_{23}\, \left(\hat{a}_{2}^{\dagger}\hat{A}_{23}+\hat{a}_{2}\hat{A}_{32}\right) \, .
\label{eq:Paper-001}
\end{eqnarray}
The first term indicates the corresponding energies of the effective three level system $E_{k}\equiv\hbar\omega_{k}$
with $k=1,2,3$ and without loosing generality establish the condition $\omega_{1}<\omega_{2}<\omega_{3}$. By denoting the one-atom states by the ket $| k_A \rangle$ with $k=1,2,3$, it is immediate to find the action of the operators $\hat{A}_{ij}$, 
\begin{equation}
\hat{A}_{ij}\,\vert k_A \rangle=\delta_{j,k}\,\vert i_{A} \rangle \,,
\end{equation}
that is,  annihilates an atom in the level $k$ and creates an atom in the level $i$. The second term establishes the Hamiltonian of the two modes of the electromagnetic field, where $\hat{n}_s$ denotes the number of photons of the mode with frequency $\Omega_s$, with $s=1,2$. The last two terms are associated to the dipolar interaction between the atom and the two-modes of the electromagnetic field. The operators ($\hat{a}_s, \hat{a}^\dagger_s$) indicate the corresponding annihilation and creation photon operators, whose commutation properties different from zero are
\begin{equation*}
[\hat{a}_s, \hat{a}^\dagger_{s^\prime}] =\delta_{s,s^\prime} \, .
\end{equation*}
From the algebraic point of view, each one of the creation and annihilation operators satisfy a Heisenberg-Weyl algebra while the operators $\hat{A}_{ij}$ with $i,j=1,2,3$ satisfy the commutation relations of a unitary algebra in three dimensions~u(3).

A straightforward calculation allows to determine that the operators, 
\begin{equation}
\hat{M}_{1}=\hat{n}_{1}+\hat{n}_{2}+\hat{A}_{33},\quad\hat{M}_{2}=\hat{n}_{2}+\hat{A}_{11}+\hat{A}_{33},\quad\hat{N}_{a}=\hat{A}_{11}+\hat{A}_{22}+\hat{A}_{33} \, , \label{eq:constant_motion}
\end{equation}
or any linear combination of them are constants of the motion of the Hamiltonian~(\ref{eq:Paper-001}).  The operator $\hat{M}_1$ denotes the total number of excitations and $N_a$ indicates the total number of particles of the system.  In this work, one has $N_a=1$.  Since we are considering dipolar transition between the atomic levels, one has that $\left|1_{A}\right\rangle $ and $\left|2_{A}\right\rangle $ have the same parity and then the dipolar transition between them is forbidden. The allowed transitions in the system are indicated by $\left|1_{A}\right\rangle \longleftrightarrow\left|3_{A}\right\rangle $
and $\left|2_{A}\right\rangle \longleftrightarrow\left|3_{A}\right\rangle $. For each one of the allowed transitions we consider photons with frequencies
$\Omega_{1}$ and $\Omega_{2}$ (Figure \ref{fig:3:level_model}).

Despite the simplicity of Hamiltonian \ref{eq:Paper-001}, it can be experimentally realizable. In fact, in the last years, many experiments have been developing this technology \cite{Blinov2004, Wilk2007, Specht2011, Ediss10030}. Roughly speaking, the cavity's length determines the possibility of coupling both modes of the field with the atom. The resonant mode frequency of the cavity with the atom is given by $f=c/2L$, where $L$ is the cavity's length. For example for $^{87}$ Rb the length needed to couple both atomic transitions is around 2.2 cm. 
 
Our model considers two field modes that are spatially independent and can be generated experimentally in two different configurations. In the first configuration, the polarization of light can be controlled independently with a double mirror cavity. The second realization consists of a single cavity with the correct length and with two orthogonal polarizations to favor the atomic lambda transitions. 

The basis to diagonalize the Hamiltonian is constituted by the tensorial
product $\left|n_{1}, \, n_{2}\right\rangle \otimes\left|k_{A}\right\rangle ,$
where $\left|n_{1}, \, n_{2}\right\rangle$ are Fock states associated to the photon number operators of modes $\Omega_{1}$
and $\Omega_{2}$, and $\left|k_{A}\right\rangle$, with $k=1,2,3,$ indicate the effective levels of the atom.
To construct the basis of the system we can act the coupling interaction terms of the Hamiltonian over the states: $\left|n_{1},n_{2};1_{A}\right\rangle ,\left|n_{1},n_{2};2_{A}\right\rangle ,\left|n_{1},n_{2};3_{A}\right\rangle$.
\begin{enumerate}
\item For $\left|n_{1},n_{2};1_{A}\right\rangle $, one gets the states
\begin{eqnarray}
\hat{a_{1}}\hat{A}_{31}\left|n_{1},n_{2};1_{A}\right\rangle  & \approx & \left|n_{1}-1,n_{2};3_{A}\right\rangle ,\qquad\hat{a_{1}}^{\dagger}\hat{A}_{13}\left|n_{1}-1,n_{2};3_{A}\right\rangle \approx\left|n_{1},n_{2};1_{A}\right\rangle \nonumber \\
\hat{a_{2}}^{\dagger}\hat{A}_{23}\left|n_{1}-1,n_{2};3_{A}\right\rangle  & \approx & \left|n_{1}-1,n_{2}+1;2_{A}\right\rangle ,\qquad\hat{a_{2}}\hat{A}_{32}\left|n_{1}-1,n_{2}+1;2_{A}\right\rangle \approx\left|n_{1}-1,n_{2};3_{A}\right\rangle 
\end{eqnarray}
\item For $\left|n_{1},n_{2};2_{A}\right\rangle $, the connected states are
\begin{eqnarray}
\hat{a}_{2}\hat{A}_{32}\left|n_{1},n_{2};2_{A}\right\rangle  & \approx & \left|n_{1},n_{2}-1;3_{A}\right\rangle ,\qquad\hat{a}_{2}^{\dagger}\hat{A}_{23}\left|n_{1},n_{2}-1;3_{A}\right\rangle \approx\left|n_{1},n_{2};2_{A}\right\rangle \nonumber \\
\hat{a}_{1}^{\dagger}\hat{A}_{13}\left|n_{1},n_{2}-1;3_{A}\right\rangle  & \approx & \left|n_{1}+1,n_{2}-1;1_{A}\right\rangle ,\qquad\hat{a}_{1}\hat{A}_{31}\left|n_{1}+1,n_{2}-1;3_{A}\right\rangle \approx\left|n_{1},n_{2}-1;3_{A}\right\rangle 
\end{eqnarray}
\item For $\left|n_{1},n_{2};3_{A}\right\rangle $, one has
\begin{eqnarray}
\hat{a_{1}}^{\dagger}\hat{A}_{13}\left|n_{1},n_{2};3_{A}\right\rangle  & \approx & \left|n_{1}+1,n_{2};1_{A}\right\rangle ,\qquad\hat{a_{1}}\hat{A}_{31}\left|n_{1}+1,n_{2};1_{A}\right\rangle \approx\left|n_{1},n_{2};3_{A}\right\rangle \nonumber \\
\hat{a_{2}}^{\dagger}\hat{A}_{23}\left|n_{1},n_{2};3_{A}\right\rangle  & \approx & \left|n_{1},n_{2}+1;2_{A}\right\rangle ,\qquad\hat{a_{2}}\hat{A}_{32}\left|n_{1},n_{2}+1;2_{A}\right\rangle \approx\left|n_{1},n_{2};3_{A}\right\rangle 
\end{eqnarray}
\end{enumerate}
Thus we have nine different states given by
\begin{eqnarray}
|\chi_{1} \rangle  & \equiv & \left|n_{1},n_{2}\right\rangle \otimes\left|1_{A}\right\rangle ,\hspace{1.6cm} |\chi_{2} \rangle \equiv\left|n_{1}-1,n_{2}+1\right\rangle \otimes\left|2_{A}\right\rangle ,\quad\left|\chi_{3} \right\rangle \equiv\left|n_{1}-1,n_{2}\right\rangle \otimes\left|3_{A}\right\rangle \, , \nonumber \\
\left|\chi_{4}\right\rangle  & \equiv & \left|n_{1}+1,n_{2}-1\right\rangle \otimes\left|1_{A}\right\rangle ,\quad\left|\chi_{5}\right\rangle \equiv\left|n_{1},n_{2}\right\rangle \otimes\left|2_{A}\right\rangle ,\hspace{1.6cm}\left|\chi_{6}\right\rangle \equiv\left|n_{1},n_{2}-1\right\rangle \otimes\left|3_{A}\right\rangle \, , \nonumber \\
\left|\chi_{7}\right\rangle  & \equiv & \left|n_{1}+1,n_{2}\right\rangle \otimes\left|1_{A}\right\rangle ,\hspace{1cm}\left|\chi_{8}\right\rangle \equiv\left|n_{1},n_{2}+1\right\rangle \otimes\left|2_{A}\right\rangle ,\hspace{1cm}\left|\chi_{9}\right\rangle \equiv\left|n_{1},n_{2}\right\rangle \otimes\left|3_{A}\right\rangle \, ,\label{eq:basis_state}
\end{eqnarray}
where in the kets $|\chi_k\rangle$, to simplify the notation we omitted the dependence on the quantum numbers $n_1$ and $n_2$ of the Fock states. Notice that, they are not independent sets of states. One can obtain the set of states of the second row from the first one by changing $n_1 \to n_1 +1$ and $n_2 \to n_2-1$. Similarly one gets the third row of states by changing $n_1 \to n_1 +1$. Therefore the Hamiltonian can only connect three different states, and thus we take the basis
\begin{equation}
\left|\chi_{1}\right\rangle \equiv\left|n_{1},n_{2}\right\rangle \otimes\left|1_{A}\right\rangle ,\quad\left|\chi_{2}\right\rangle \equiv\left|n_{1}-1,n_{2}+1\right\rangle \otimes\left|2_{A}\right\rangle ,\quad\left|\chi_{3}\right\rangle \equiv\left|n_{1}-1,n_{2}\right\rangle \otimes\left|3_{A}\right\rangle \, .
\label{old}
\end{equation}
Therefore the matrix Hamiltonian is given by
\[
\hat{H}=\left(\begin{array}{ccc}
E_{\left|1\right\rangle } & 0 & \hbar\mu_{13}\sqrt{n_{1}}\\
0 & E_{\left|2\right\rangle } & \hbar\mu_{23}\sqrt{n_{2}+1}\\
\hbar\mu_{13}\sqrt{n_{1}} & \hbar\mu_{23}\sqrt{n_{2}+1} & E_{\left|3\right\rangle }
\end{array}\right),
\]
where the energies of the basis states are denoted by
\begin{eqnarray*}
E_{\left|1\right\rangle }  =  \hbar\left(\omega_{1}+\Omega_{1}n_{1}+\Omega_{2}n_{2}\right)\, , \quad 
E_{\left|2\right\rangle }  =  \hbar\left(\omega_{1}+\Omega_{1}\left(n_{1}-1\right)+\Omega_{2}\left(n_{2}+1\right)\right)\, , \quad 
E_{\left|3\right\rangle }  =  \hbar\left(\omega_{1}+\Omega_{1}\left(n_{1}-1\right)+\Omega_{2}n_{2}\right).
\end{eqnarray*}
It is convenient to rewrite the Hamiltonian in terms of experimental
parameters that can be precisely controlled, that is, the frequencies of
the two photon modes and the detuning parameters of the transitions. Thus, they satisfy the following relations,
\begin{equation}
\omega_{31}=\Omega_{1}+\Delta_{13},\quad\omega_{32}=\Omega_{2}+\Delta_{23},\quad\omega_{21}=\omega_{31}-\omega_{32}=\Omega_{1}-\Omega_{2}+\Delta_{13}-\Delta_{23},\quad\omega_{31}=\omega_{3}-\omega_{1},\quad\omega_{32}=\omega_{3}-\omega_{2}.\label{eq:Paper-011}
\end{equation}
Introducing (\ref{eq:Paper-011}), the Hamiltonian can be rewritten in the form
\begin{eqnarray}
\hat{H} & = & \hbar\left(\omega_{1}+\Omega_{1}n_{1}+\Omega_{2}n_{2}\right) \, \hat{I}_{3}\nonumber \\
 & + & \hbar\left(\begin{array}{ccc}
0 & 0 & \mu_{13}\sqrt{n_{1}}\\
0 & \Delta_{13}-\Delta_{23} & \mu_{23}\sqrt{n_{2}+1}\\
\mu_{13}\sqrt{n_{1}} & \mu_{23}\sqrt{n_{2}+1} & \Delta_{13}
\end{array}\right) \, . \label{eq:Paper-012}
\end{eqnarray}

By taking the action of the constants of motion $\hat{M}_{1}$ and $\hat{M}_{2}$ on the basis states~(\ref{old}), one gets
\begin{eqnarray*}
\hat{M}_{1} \left|\chi_{k}\right\rangle  & = & \left(n_{1}+n_{2}\right) \left|\chi_{k}\right\rangle ,
\qquad\hat{M}_{2} \left|\chi_{k}\right\rangle =\left(n_{2}+1\right) \left|\chi_{k} \right\rangle \, .
\end{eqnarray*}
with $k=1,2,3$.  Thus one can make the replacement of the number of photons $n_{1},n_{2}$
in terms of the eigenvalues $m_k$ of the constants of the motion $\hat{M}_{k}$, with $k=1,2$.
one has
\begin{equation}
\vert\chi_{1}\rangle=\vert m_{1}-m_{2}+1\,,m_{2}-1;1_{A}\rangle\,, \quad \vert\chi_{2}\rangle=\vert m_{1}-m_{2}\,,m_{2};2_{A}\rangle\,, \quad \vert\chi_{3}\rangle=\vert m_{1}-m_{2}\,,m_{2}-1;3_{A}\rangle.\label{eq:Paper-009}
\end{equation}
In terms of the corresponding eigenvalues $m_k$ of the excitation operators, the matrix Hamiltonian takes the form, 
\begin{eqnarray}
\hat{H} & = & \hbar\left[\omega_{1}+\Omega_{1}\,(m_{1}-m_{2}+1)+\Omega_{2}\,(m_{2}-1)\right]\hat{I}_{3}\nonumber \\
 & + & \left(\begin{array}{ccc}
0 & 0 & \mu_{13}\,\sqrt{m_{1}-m_{2}+1}\\
0 & \Delta_{13}-\Delta_{23} & \mu_{23}\,\sqrt{m_{2}}\\
\mu_{13}\,\sqrt{m_{1}-m_{2}+1} & \mu_{23}\,\sqrt{m_{2}} & \Delta_{13}
\end{array}\right).
\label{eq:Paper-013}
\end{eqnarray}
The eigenvalues $m_{1}$ and $m_{2}$ of the excitation operators $\hat{M}_1$ and $\hat{M}_2$, respectively are
restricted to the following intervals: $0\leq m_{1}\leq M_{0}$
and $0\leq m_{2}\leq m_{1}+1$, where $M_{0}\equiv n_{1max}+n_{2max}+1$.
Moreover, matrices (\ref{eq:Paper-012}) and (\ref{eq:Paper-013})
are diagonal for specific states. Therefore, exists a set of states
for which there is no coupling with the field. These states are known
as dark states and are discussed in the next section. 

\section{Dressed States}
The diagonalization of the matrix Hamiltonian~(\ref{eq:Paper-013}) can be obtained in analytical because involves the solution of third order algebraic equation. However a simple solution can be obtained when the detuning condition $\Delta_{13}=\Delta_{23}$ is taken.  For this case, the energy eigenvalues are given by
\begin{equation}
E_{0}\left(m_{1},m_{2}\right)=\hbar\left[\omega_{1}+\Omega_{1}\left(m_{1}-m_{2}+1\right)+\Omega_{2}\left(m_{2}-1\right)\right] \, , \quad 
E_{\pm}\left(m_{1},m_{2}\right)=E_{0}\left(m_{1},m_{2}\right)+\hbar\frac{\Delta_{13}}{2}\pm\hbar\varepsilon_{2} \, ,
\label{eq:Paper-014}
\end{equation}
where we define 
\begin{equation*}
\varepsilon_{2}\equiv\sqrt{\left(\frac{\Delta_{13}}{2}\right)^{2}+(m_{1}-m_{2}+1)\,\mu_{13}^{2}+m_{2}\,\mu_{23}^{2}.}
\end{equation*}
The corresponding eigenvectors are 
\begin{eqnarray}
\left|\Psi_{0}\left(m_{1},m_{2}\right)\right\rangle  & = & \frac{1}{\sqrt{\varepsilon_{2}^{2}-\frac{\Delta_{13}^{2}}{4}}}\left\{ -\mu_{23}\,\sqrt{m_{2}}\,\left|\chi_1\right\rangle +\mu_{13}\,\sqrt{m_{1}-m_{2}+1}\,\left| \chi_2\right\rangle \right\} \,,\label{eq:Paper-016} \\
\left|\Psi_{\pm}\left(m_{1},m_{2}\right)\right\rangle  & = & \frac{1}{\sqrt{2\,\varepsilon_{2}^{2}\pm\Delta_{13}\,\varepsilon_{2}}}\left\{ \mu_{13}\,\sqrt{m_{1}-m_{2}+1}\,\left| \chi_1\right\rangle +\mu_{23}\,\sqrt{m_{2}}\,\left|\chi_2\right\rangle +\left[\frac{\Delta_{13}}{2}\pm\varepsilon_{2}\right]\,\left|\chi_3\right\rangle \right\} ,
\label{eq:Paper-017}
\end{eqnarray}
which are written in the basis (\ref{eq:Paper-009}). State (\ref{eq:Paper-016})
is known as a dark state, because the expectation value of the two coupling terms of the Hamiltonian with respect to $\left|\Psi_{0}\right\rangle $ are zero, that is,
\begin{equation*}
\left\langle \Psi_{0}\left(m_{1},m_{2}\right)\right|\hat{a_{1}}\hat{A}_{31}+\hat{a_{1}}^{\dagger}\hat{A}_{13}\left|\Psi_{0}\left(m_{1},m_{2}\right)\right\rangle = \left\langle \Psi_{0}\left(m_{1},m_{2}\right)\right|\hat{a_{2}}\hat{A}_{32}+\hat{a_{2}}^{\dagger}\hat{A}_{23}\left|\Psi_{0}\left(m_{1},m_{2}\right)\right\rangle   =  0 \, .
\end{equation*}
This is 
reflected also in its corresponding energy eigenvalue (\ref{eq:Paper-013}), which is composed by
the sum of the lowest atomic state energy and the energies of the
two modes of the field. States (\ref{eq:Paper-017}) 
are known as brilliant states, due to mean value of the coupling terms
of the Hamiltonian in $\left|\Psi_{-}\right\rangle $ and $\left|\Psi_{+}\right\rangle $, which 
depend linearly on the Rabi frequency and the number of excitations,
\begin{eqnarray}
\left\langle \Psi_{\pm}\left(m_{1},m_{2}\right)\right|\hat{a_{1}}\hat{A}_{31}+\hat{a_{1}}^{\dagger}\hat{A}_{13}\left|\Psi_{\pm}\left(m_{1},m_{2}\right)\right\rangle  & = & \pm \, \frac{\mu_{13}}{\varepsilon_{2}}\left(m_{1}-m_{2}+1\right) \, , \nonumber \\
\left\langle \Psi_{\pm}\left(m_{1},m_{2}\right)\right|\hat{a_{2}}\hat{A}_{32}+\hat{a_{2}}^{\dagger}\hat{A}_{23}\left|\Psi_{\pm}\left(m_{1},m_{2}\right)\right\rangle  & = & \pm \,  \frac{\mu_{23}}{\varepsilon_{2}}m_{2}  \, .
\nonumber
\end{eqnarray}
On the other hand, it is important to know that there are other states that are unaffected by the matter-field
coupling interaction. These states are also known as dark states and they have zero photons in the mode 1 and/or zero photons in the second mode of the electromagnetic field in the cavity,
\begin{equation}
|\psi_{d_1}(n_1)\rangle=| 0, n_2;1_A \rangle, \, \quad |\psi_{d_2}(n_2)\rangle=\left|n_{1},0;2_{A}\right\rangle \, ,
\label{eq:darkstates}
\end{equation}
whose corresponding energies are,
\[
\hat{H}|0,n_2;1_{A}\rangle =\hbar (\omega_{1} + \Omega_2\, n_2) | 0,n_2;1_{A}\rangle \,   , \quad 
\hat{H}\left|n_{1},0;2_{A}\right\rangle =\hbar ( \omega_{2}+ \Omega_{1} \, {n}_{1}) \left|n_{1},0;2_{A}\right\rangle \, .
\]
Thus, its time evolution is given by 
\[
\hat{H}|0,n_2;1_{A}\rangle_t = \mathrm{e}^{-i (\omega_{1} + \Omega_2 \, n_2)\, t } |0,n_2;1_{A}\rangle , \quad  |n_{1},0;2_{A}\rangle_t =\mathrm{e}^{-i\left(\omega_{2}+\Omega_{1}n_{1}\right)t}\left|n_{1},0;2_{A}\right\rangle  \,  .
\]
Although the condition $\Delta_{13}-\Delta_{23}=0$ simplifies the form of the analytic solution of 
the eigensystem, the general case $\Delta_{13}-\Delta_{23}\neq0$ can also be calculated. One finds a similar structure of the energy levels, the main difference is that the $E_{0}$ eigenvalue depends of the coupling strengths $\mu_{13}$ and $\mu_{23}$. Depending on the sign of the difference $\Delta_{13}-\Delta_{23},$
$E_{0}$ approaches to the new eigenvalues $E_{+}$ or $E_{-}$, for certain values of the Rabi frequencies. In consequence, one finds that the corresponding eigenstate to $E_0$ is not a dark state. 

\section{Time Evolution Operator}

The evolution operator of the Hamiltonian system can be obtained analytically in several forms. Here we use that the dressed states basis  are connected with the basis states~(\ref{eq:Paper-009}) through an orthogonal
transformation $\hat{\cal{O}}$, here we are using the condition of equal detunings,  
\begin{equation}
\left(\begin{array}{c}
\left|\Psi_{+}(m_1,m_2)\right\rangle\\
\left|\Psi_{0} (m_1,m_2)\right\rangle \\
\left|\Psi_{-} (m_1,m_2)\right\rangle 
\end{array}\right)
=\hat{\cal{O}}\left(\begin{array}{c}
\left|\chi_{1}\right\rangle \\
\left|\chi_{2}\right\rangle \\
\left|\chi_{3}\right\rangle 
\end{array}\right) .
\label{eqnonew1}
\end{equation}
Considering that a general wave packet can be expanded into the basis states~(\ref{eq:Paper-009}), that is,
\begin{equation}
| \psi(0)\rangle = \sum_{m_1,m_2, k} \, c (m_1,m_2,k) \, |\chi_k (m_1,m_2)\rangle \, ,
\label{psizero}
\end{equation}
where we put explicitly the dependence on the quantum numbers associated to the operators $\hat{M}_1$ and $\hat{M}_2$. Thus, one has a wave packet constructed by the product of Fock states associated to the modes of the electromagnetic field in the cavity times the state of the three level atom, which in general is not an eigenstate of the Hamiltonian model. For this reason, one finds convenient to multiply the expression~(\ref{eqnonew1}) by the transpose, $\hat{\cal{O}}^T$,  that is, 
\begin{equation}
\hat{\cal{O}}^T=\left(\begin{array}{ccc}
\frac{\mu_{13}\sqrt{m_{1}-m_{2}+1}}{\sqrt{2\,\varepsilon_{2}^{2}+\Delta_{13}\,\varepsilon_{2}}} & \frac{-\mu_{23}\,\sqrt{m_{2}}}{\sqrt{\varepsilon_{2}^{2}-\frac{\Delta_{13}^{2}}{4}}} & \frac{\mu_{13}\,\sqrt{m_{1}-m_{2}+1}}{\sqrt{2\,\varepsilon_{2}^{2}-\Delta_{13}\,\varepsilon_{2}}}\\
\frac{\mu_{23}\sqrt{m_{2}}}{\sqrt{2\,\varepsilon_{2}^{2}+\Delta_{13}\,\varepsilon_{2}}} & \frac{\mu_{13}\,\sqrt{m_{1}-m_{2}+1}}{\sqrt{\varepsilon_{2}^{2}-\frac{\Delta_{13}^{2}}{4}}} & \frac{\mu_{23}\,\sqrt{m_{2}}}{\sqrt{2\,\varepsilon_{2}^{2}-\Delta_{13}\,\varepsilon_{2}}}\\
\frac{1}{2\,\varepsilon_{2}}\sqrt{2\,\varepsilon_{2}^{2}+\Delta_{13}\,\varepsilon_{2}} & 0 & -\frac{1}{2\,\varepsilon_{2}}\sqrt{2\,\varepsilon_{2}^{2}-\Delta_{13}\,\varepsilon_{2}}
\end{array}\right).\label{eq:Paper-022}
\end{equation}
Thus the basis states can be written as linear combinations of the dressed states, and since the Hamiltoniano (\ref{eq:Paper-001}) is diagonal in the dressed states basis, the action of the time evolution operator
is straightforward, that is,
\begin{equation}
\left(\begin{array}{c}
|\chi_{1},t \rangle \\
|\chi_{2},t \rangle \\
|\chi_{3},t\rangle 
\end{array}\right)=\hat{\cal{O}}^T\hat{D}(t) \, \hat{\cal{O}} \, \left(\begin{array}{c}
|\chi_{1} \rangle \\
|\chi_{2} \rangle \\
|\chi_{3}\rangle 
\end{array}\right),\label{eq:Paper-023}
\end{equation}
where the matrix $\hat{D}(t)$ is a diagonal matrix of the form,
\[
\hat{D}(t)=\left\{ \mathrm{e}^{-i\frac{E_{+}}{\hbar}t},\mathrm{e}^{-i \frac{E_{0}}{\hbar}t},\mathrm{e}^{-i\frac{E_{-}}{\hbar}t}\right\} .
\]
Therefore, the time evolution operator can be defined as athree dimensionsl matrix of the form
\begin{eqnarray}
\hat{U}\left(t\right) &=&\left(
\begin{array}{ccc}
 u_{11}\left(m_{1},m_{2},t\right) & u_{12}\left(m_{1},m_{2},t\right)  & u_{13}\left(m_{1},m_{2},t\right)  \\
 u_{21}\left(m_{1},m_{2},t\right) & u_{22}\left(m_{1},m_{2},t\right)  & u_{23}\left(m_{1},m_{2},t\right)  \\
 u_{31}\left(m_{1},m_{2},t\right) & u_{32}\left(m_{1},m_{2},t\right)  & u_{33}\left(m_{1},m_{2},t\right)
\end{array}
\right) \, ,
\label{matU}
\end{eqnarray}
where $u_{ij}\left(m_{1},\,m_{2},\,t\right)$, with $i,j=1,2,3$ are given explicitely in an appendix. 

\subsection{Initial Wave Packet}
We consider an initial wave function $\vert \psi_0\rangle$ which lives in the Hilbert space ${\cal H}$ constituted by the tensorial product of the Hilbert space of the field ${\cal H}_F$ times de Hilbert space for the matter ${\cal H}_M$, that is, ${\cal H}  ={\cal H}_F \otimes {\cal H}_M$. If we are considering $N_a$ atoms for the matter sector, the Hilbert space ${\cal H}_M$  has $(N_a+1)(N_a+2)/2$ dimensions while for the two-modes of the electromagnetic field we consider a maximum number of photons $(\nu_{1},\nu_{2})$. Thus the dimension of the Hilbert space of the system is given by $d_{\cal H} = (N_a+1)\, (N_a+2) (\nu_{1}+1)(\nu_{2} +1)/2$. 

Therefore the initial state can be written in the separable form $\vert \psi_0\rangle= |\Phi_F \rangle\otimes | \chi_M\rangle$,  where for the field sector one writes
\begin{equation}
|\Phi_F \rangle\equiv | \phi(\nu_{1},\, \nu_{2})\rangle= \sum^{\nu_{1}}_{n_1=0} \, \sum^{\nu_{2}}_{n_2=0} \, C_{n_1,n_2} \, | n_1, n_2\rangle \, ,
\end{equation}
where $| n_s\rangle$, with $s=1,2$ denotes a Fock state for the mode $s$ of the field.  The matter part can be written as a coherent state of unitary group in three dimensions U$(3)$, belonging to a totally symmetric representation with $N_a$ particles, that is,
\begin{equation}
| \chi\rangle_M\equiv|N_{a};\hat{\gamma}_{k}\rangle =\frac{1}{\sqrt{N_{a}!}}\frac{\left(\gamma_{1}\,\hat{b}_{1}^{\dagger}+\gamma_{2}\, \hat{b}_{2}^{\dagger}+\gamma_{3}\, \hat{b}_{3}^{\dagger}\right)^{N_{a}}}{\left(\left|\gamma_{1}\right|^{2}+\left|\gamma_{2}\right|^{2}+\left|\gamma_{3}\right|^{2}\right)^{N_{a}/2}}|0\rangle _{M}\,. 
\end{equation}
where $|0\rangle _{M}=|0 \, 0\, 0\rangle$ is the vacuum state of matter, and $b_{k}^{\dagger}$ are boson creation operators with $\gamma_{k}$, with $k=1,2,3$, in general, denoting complex parameters. 

Since in this work, we are considering a system with one atom, $N_{a}=1$, for the matter sector and the product of two Glauber coherent states, one for each mode of the electromagnetic field $|\alpha_s\rangle$, with $s=1,2$. The separable initial state can be written in the form 
\begin{eqnarray}
|\psi_0\rangle &=& | \alpha_1 \alpha_2 ; \hat{\gamma}_1\,\hat{\gamma}_2\,\hat{\gamma}_3 \rangle  \nonumber 
\\
&=& \sum_{n_1, \, n_2, \, k} \,   e^{-(|\alpha_1|^2/2 + |\alpha_2|^2/2)} \,  \frac{(\alpha_1)^{n_1}}{\sqrt{n_1!}} \, \frac{(\alpha_2)^{n_2}}{\sqrt{n_2!}} \, \hat{\gamma}_k \, |n_1 \, n_2 ; k_A\rangle \, ,
\label{expansion}
\end{eqnarray}
with the complex parameters  $\hat{\gamma}_{k}=\zeta_{k}\, e^{i \, \theta_{k}}$ defined by
\begin{equation*}
\hat{\gamma}_{k}=\frac{\zeta_{k} \, e^{i \, \theta_{k}} }{\sqrt{\zeta_{1}^{2}+\zeta_{2}^{2}+\zeta_{3}^{2}}} \, ,
\end{equation*}
so that the normalization condition $\sum_{k=1}^{3}\left|\hat{\gamma}_{k}\right|^{2}=1$ is hold. 

Using the constants of motion of Hamiltonian system, associated to the number of excitations $\hat{M}_1$ and $\hat{M}_2$, the kets of the expansion~(\ref{expansion}) can be replaced by
\begin{equation}
| n_1 \, n_2; k_A\rangle \to | m_1-m_2 + \delta_{k\, 1} , \, m_2-\delta_{k\, 1} -\delta_{k\, 3} ; k_A \rangle \, ,
\end{equation}
where the quantum number $m_1$($m_2$) denotes the eigenvalue of $\hat{M}_1$($\hat{M}_2$). Besides for a maximum value $M_0$ of the total number of excitations, the range of these eigenvalues is the following
\begin{equation}
0\leq m_1\leq M_0 , \quad \delta_{k\,1} +\delta_{k\,3}\leq m_2 \leq m_1 + \delta_{k\,1} \, ,
\end{equation}
which of course depends on the state occupied by the atom. Thus, one concludes that the dimension of the Hilbert space of the system is given by $d_{\cal H}= (M_0+1)(3\, M_0 +4)/2$.

For the two-modes of the electromagnetic field in the cavity, one has Glauber coherent states $|\alpha_s\rangle$, which constitute an overcomplete basis, where the photons has a Poissonian distribution funtion, that is
\begin{equation}
P(\bar{n}, k) = e^{-\bar{n}} \, \frac{\bar{n}^k}{k!} \, ,
\end{equation}
where the mean value of the number of photons is given by $\bar{n} =|\alpha_s|^2$, and the corresponding standard deviation $\Delta n=\bar{n}$; implying that the corresponding Mandel number $Q_M=0$. This number consitutes an alternative measure of the second order photon statistical properties of a beam light.

In phase space the Husimi Q function is defined by
\[
{\cal Q}_H = \frac{1}{\pi} \, |\langle \beta |\alpha\rangle|^2 = \frac{1}{\pi} \, e^{-|\alpha - \beta|^2}  \, ,
\]
where $\beta$ is a complex number. Here it is convenient to define the quadrature operators of a quantized electromagnetic field of frequency $\omega$,
\begin{equation}
\hat{X} = \sqrt{\frac{m\, \omega}{2\, \hbar}} \hat{q} = \frac{1}{2} (\hat{a}^\dagger + \hat{a}) \, , \quad \hat{Y} = \sqrt{\frac{1}{2\, m\, \hbar\, \omega}} \, \hat{p} = \frac{i}{2} (\hat{a}^\dagger - \hat{a}) \, ,
\end{equation}
which satisfy the commutation relations $[\hat{X},\hat{Y}] =\frac{i}{2}$, which leads to a dimensionless form of  Heisenberg position-momentum uncertainty relation $(\Delta X)^2 (\Delta Y)^2\geq \frac{1}{16}$. Then we consider $\beta=X + i Y$. The Husimi function is always positive and it is bounded ${\cal Q}_H \leq \frac{1}{\pi}$.  

The moments of the Husimi function of quantum states are good measures of complexity \cite{Sugita2003}, in particular their delocalization property is a sign of entanglement. Thus we determine the area occupied for the Husimi function of a coherent state, in phase space. It is given by the inverse expression of the second moment of the Husimi function,
\begin{equation}
{\cal M}^{(2)} = \pi \, \int^{\infty}_{-\infty} d Y \int^\infty_{-\infty} d X \, {\cal Q}^2_H(X,Y) =1 \, . \label{second_momentum}
\end{equation}
Thus ${\cal A} = 1/{\cal M}^{(2)} =1$ is the minimum area in phase space which can be occupied by a coherent state.  Later on this section we are going to determine what happen with the area of the Husimi function in phase space at time $t$. 

For the matter sector, it is convenient to construct the three dimensional reduced density matrix, 
\begin{equation*}
\hat{\rho}_{A}\left(0\right)=\sum_{i,j=1}^{3}\hat{\gamma}_{i}\hat{\gamma}_{j}^{*}\left|i_{A}\right\rangle \left\langle j_{A}\right|\equiv\left(\begin{array}{ccc}
\left|\hat{\gamma}_{1}\right|^{2} & \hat{\gamma}_{1}\hat{\gamma}_{2}^{*} & \hat{\gamma}_{1}\hat{\gamma}_{3}^{*}\\
\hat{\gamma}_{2}\hat{\gamma}_{1}^{*} & \left|\hat{\gamma}_{2}\right|^{2} & \hat{\gamma}_{2}\hat{\gamma}_{3}^{*}\\
\hat{\gamma}_{3}\hat{\gamma}_{1}^{*} & \hat{\gamma}_{3}\hat{\gamma}_{2}^{*} & \left|\hat{\gamma}_{3}\right|^{2}
\end{array}\right) \, .
\end{equation*}
The diagonal elements $\left[\hat{\rho}_A(0)\right]_{kk}$ indicates the occupation probability of the state $\ket{k_A}$, with $k=1,2,3$. Thus, condition $\mathrm{Tr}\left[\hat{\rho}_A(0)\right]=1$ is satisfied. The non-diagonal terms allows us calculate the coherence, 
\begin{equation}
C_{\rm coh} = 2 |\hat{\gamma}_1 \, \hat{\gamma}^*_2| +2 |\hat{\gamma}_1 \, \hat{\gamma}^*_3|  + 2 |\hat{\gamma}_2 \, \hat{\gamma}^*_3| \, .
\end{equation}

Now we calculate the expectation values of the constants of motion $\hat{H}$, $\hat{M}_1$, and $\hat{M}_2$ with respect to the initial wave packet of expression~(\ref{expansion}). The total system energy is given by $E=\braket{\hat{H}}_t$, thus it only depends on the parameters that define the initial state, 
\begin{eqnarray}
E &=& \sum_k \, \omega_k \, |  \hat{\gamma}_k| ^2 +\sum_s \Omega_s | \alpha_s|^2 
- \mu_{13} (\alpha^*_1 \, \hat{\gamma}^*_1 \, \hat{\gamma}_3 + \alpha_1 \, \hat{\gamma}_1 \, \hat{\gamma}^*_3) - \mu_{23} (\alpha^*_2 \, \hat{\gamma}^*_2 \, \hat{\gamma}_3 + \alpha_2 \, \hat{\gamma}_2 \, \hat{\gamma}^*_3)  \, .
\end{eqnarray}
While for the excitation number operators one has
\begin{equation}
\langle \hat{M}_1 \rangle_0 = |\alpha_1|^2 + |\alpha_2|^2 + |\hat{\gamma}_3|^2 \, , \quad \langle \hat{M}_2 \rangle_0 = |\alpha_2|^2 + |\hat{\gamma}_1|^2 +  |\hat{\gamma}_3|^2 \, .
\end{equation}
Notice that the Hamiltonian operator can not change the eigenvalues $(m_1,m_2)$ of the mentioned constants of motion $\hat{M}_1$ and $\hat{M}_2$, respectively.
Then it is immediate and convenient to determine the corresponding joint probability, 
\begin{eqnarray}
{\cal P}(m_1,m_2,t) &=& e^{-|\alpha_1|^2 -|\alpha_2|^2 } (|\alpha_1|^2)^{m_1-m_2} \, (|\alpha_2|^2)^{m_2-1} \nonumber \\
& \times& \left( \frac{m_2 \, |\alpha_1|^2 \, |\hat{\gamma}_1|^2 + (m_1-m_2 +1)\, |\alpha_2|^2 \, |\hat{\gamma}_2|^2 + m_2( m_1-m_2+1) \, |\hat{\gamma}_3|^2 } {\Gamma(m_1-m_2+2) \, \Gamma(m_2+1)} \right)\, ,
\label{probM1M2}
\end{eqnarray}
which is of course valid for any time $t$.  

The probability of having $m_1$ excitations is given by
\begin{equation}
{\cal P}_{M_1}(m_1,t) =  e^{-|\alpha_1|^2 -|\alpha_2|^2 }   \left\{ \frac{(|\alpha_1|^2 + |\alpha_2|^2)^{m_1}}{m_1!} (|\hat{\gamma}_1|^2 + |\hat{\gamma}_2|^2)  +  \frac{(|\alpha_1|^2 + |\alpha_2|^2)^{m_1-1}}{(m_1-1)!} 
|\hat{\gamma}_3|^2 \right\} \, ,
\end{equation}
while for the probability of getting $m_2$ excitations, one arrives to the result
\begin{equation}
{\cal P}_{M_2}(m_2,t) =  e^{-|\alpha_2|^2 }   \left\{ \frac{( |\alpha_2|^2)^{m_2}}{m_2!}  |\hat{\gamma}_2|^2  +  \frac{(|\alpha_2|^2)^{m_2-1}}{(m_2-1)!} 
( |\hat{\gamma}_1|^2 + |\hat{\gamma}_3|^2) \right\} \, .
\end{equation}
One concludes that the joint probability function of $(m_1,m_2)$ depends only on the possible eigenvalues of the constants of the motion associated to the initial state. Then the last probability distribution functions can be used to to reconstruct the initial wave packet in an experiment to study the dynamics described by the Hamiltonian~(\ref{eq:Paper-001}).

\subsection{Wave Packet at time $t$}

First we write the initial wave packet~(\ref{expansion}) in terms of the eigenvalues of constants of the motion $\hat{M}_1$ and $\hat{M}_2$, thus the initial wave packet can be written in the form,
\begin{eqnarray}
|\psi_0\rangle &=&  e^{-(|\alpha_1|^2/2 + |\alpha_2|^2/2)}  \sum_{m_1, \, m_2, \, k} \,  \,  \frac{(\alpha_1)^{m_1-m_2+\delta_{k,1}}}{\sqrt{(m_1-m_2+\delta_{k,1})!}} \, \frac{(\alpha_2)^{m_2-\delta_{k,1}-\delta_{k,3}}}{\sqrt{(m_2-\delta_{k,1}-\delta_{k,3})!}} \, \nonumber \\
&\times& \hat{\gamma}_k \, |m_1-m_2+\delta_{k,1} \, , m_2-\delta_{k,1}-\delta_{k,3} ; k_A\rangle \,  \nonumber 
\\
&\equiv& \sum_{m_1,m_2, k} \, c (\alpha_s, \hat{\gamma}_k, m_1,m_2) \, |m_1-m_2+\delta_{k,1} \, , m_2-\delta_{k,1}-\delta_{k,3} ; k_A\rangle \, ,
\label{expansion2}
\end{eqnarray}
with $s=1,2$ in the coefficient $c (\alpha_s, \hat{\gamma}_k, m_1,m_2)$. 

Now one takes the action of the evolution operator $\hat{U}(t)=\hat{\cal{O}}^T\hat{D}(t) \, \hat{\cal{O}}$, written explicitely in expression~(\ref{matU}), to determine the wave packet at an arbitrary time $t$, 
\begin{equation}\label{wavepacket}
\vert\psi\left(t\right)\rangle=\sum_{m_1,m_2,k} \, G_k (\alpha_s, \hat{\gamma}_k, m_1,m_2,t)\, |m_1-m_2+\delta_{k,1} \, , m_2-\delta_{k,1}-\delta_{k,3} ; k_A\rangle\, ,
\end{equation}
where we have defined the coefficient $G_{k}$ and it is written in terms of the matrix elements of the evolution operator as follows,
\begin{equation}
G_{k}(\alpha_s, \hat{\gamma}_k, m_1,m_2,t) = \sum^3_{k^\prime=1} \, c (\alpha_s, \hat{\gamma}_{k^\prime}, m_1,m_2) \, u_{k^\prime ,k}(m_1,m_2,t)  \, .
\end{equation}

Now we study the evolution of the expectation values of several matter and field observables. We start by considering the autocorrelation function, that is, 
\begin{eqnarray}
{\cal A}(t) &=& |\langle \psi(0) | \psi(t)\rangle|^2 \, \nonumber \\
&=& \left| \sum_{m_1,m_2,k} c^* (\alpha_s, \hat{\gamma}_{k}, m_1,m_2) \, G_{k}(\alpha_s, \hat{\gamma}_k, m_1,m_2,t) \right| ^2 \, , \label{Autocorrelation}
\end{eqnarray}
it measures the fidelity between the wave packet states $|\psi(t)\rangle$ and $|\psi_0\rangle$. The expression can be generalized to calculate the fidelity at two times.  Here we relate the dynamics of $A(t)$ to the behavior of the expectation values of atomic and field observables as it will be shown later on the results section.

\subsubsection{Reduced density matrix of the atom}
The expectation values of the field and matter observables are easily calculated by working with the reduced density matrix of the field $\hat{\rho}_F(t)$ and of the atom $\hat{\rho}_A(t)$ sectors, respectively. Furthermore, the density matrix describe completely the system, just as the wave function.  Then one calculates the density matrix of the system $\rho(t) = |\psi(t)\rangle\langle\psi(t)|$, and afterwards take the trace with respect to the field part to determine the reduced density matrix of the atom,
\begin{equation}
\rho^{(A)}_{k_1,k_2} (t) = \sum_{m_1,m_2}\,  G_{k_1}(\alpha_s, \hat{\gamma}_{k_1}, m_1,m_2,t) G^*_{k_2}(\alpha_s, \hat{\gamma}_{k_2}, m_1-\delta_{k_1,3} + \delta_{k_2,3},m_2 -\delta_{k_1, 1} -\delta_{k_1,3} + \delta_{k_2,1} +\delta_{k_2,3},t) \, , 
\end{equation}
which can be written by a three dimensional matrix,
\begin{equation}
\hat{\rho}^{A}\left(t\right)=\left(\begin{array}{ccc}
\mathcal{P}_{1}\left(t\right) & \chi_{12}\left(t\right) & \chi_{13}\left(t\right)\\
\chi_{21}\left(t\right) & \mathcal{P}_{2}\left(t\right) & \chi_{23}\left(t\right)\\
\chi_{31}\left(t\right) & \chi_{32}\left(t\right) & \mathcal{P}_{3}\left(t\right)
\end{array}\right)\,. \label{rhoA}
\end{equation}
Each one of the diagonal elements $\mathcal{P}_k(t)$, with $k=1,2,3$ indicates the probability of finding the atom in state $\ket{k_A}$ at any time $t$ and is determined by the expression,
\begin{eqnarray}
\mathcal{P}_{k}(t)&=&\sum_{m_1, m_2} \left|G_{k}\left(\alpha_s,\hat{\gamma}_k, m_{1},m_{2},t\right)\right|^{2} \, .
\label{AtomicProb}
\end{eqnarray}
Because we are considering the one atom case, the occupation probabilities of the atomic levels are also obtained by means of the expectation value of the operators $\hat{A}_{kk}$, with $k=1,2,3$, i.e., $\mathcal{P}_k(t)=\braket{\psi(t)|\hat{A}_{kk}|\psi(t)}$. Due to the natural dynamics of the system, probabilities \eqref{AtomicProb} have an associated fluctuation, equivalent to the fluctuation of $\hat{A}_{kk}$:
\begin{equation}
\left(\Delta \braket{\hat{A}_{kk}} \right)^2=\left(\Delta\mathcal{P}_{kk}\right)^{2}=\mathcal{P}_{k}\left(1-\mathcal{P}_{k}\right)\,. 
\end{equation}

The non-diagonal terms $\chi_{ij}(t)$ can be calculated in a similar form, that is, they are related to the expectation values of the atom operators, $\chi_{ij}(t)= \langle \hat{A}_{ji}\rangle_t$, with $i\neq j$. They determine the coherence properties of the atom $C_{\rm coh}(t)$ through the expression,
\begin{equation}
C_{\rm coh}(t) = 2 |\chi_{12}(t)| +2 |\chi_{13}(t)|  + 2 |\chi_{23}(t)| \, , \label{CoherenceFunction}
\end{equation}
and each one of the them is given by,
\begin{eqnarray}
\chi_{12}(t)=\chi_{21}(t)^*&=&\sum_{m_1,m_2}
G_{1}\left(\alpha_s, \hat{\gamma}_1, m_1\, , m_2,t\right) \, G_{2}^{*}\left(\alpha_s, \hat{\gamma}_2, m_{1}, m_2-1,t\right) \, ,  \nonumber \\
\chi_{13}(t)=\chi_{31}^*(t)&=&\sum_{m_1,m_2}
G_{1}\left(\alpha_s, \hat{\gamma}_1, m_1, m_2,t\right)\, G_{3}^{*}\left(\alpha_s, \hat{\gamma}_3, m_1+1,m_2,t\right) \, , \nonumber \\
\chi_{23}(t)=\chi_{32}^*(t)&=& \sum_{m_1,m_2}G_{2}\left(\alpha_s, \hat{\gamma}_2, m_1, m_2,t\right) \, G_{3}^{*}\left(\alpha_s, \hat{\gamma}_3, m_1+1, m_2+1,t\right)\, .
\end{eqnarray}
Equalities $\chi_{ji}^*=\chi_{ij}$ are satisfied because $\hat{\rho}_A$ must be a hermitian matrix. 

\subsubsection{Reduced density matrix of the electromagnetic field}

By taking the trace of the density matrix associated to the wave packet at time $t$ with respect to the atom sector one determines the reduced density matrix of the electromagnetic field in the cavity,
\begin{equation}
\rho^{(F)}_{\nu_1,\nu_2; \mu_1,\mu_2} (t) = \sum_k \, G_{k}\left(\alpha_s, \hat{\gamma}_k, \nu_1+\nu_2 +\delta_{k,3}, \nu_2 +\delta_{k, 1} + \delta_{k,3},t\right) \, G^*_{k}\left(\alpha_s, \hat{\gamma}_k, \mu_1+\mu_2 +\delta_{k,3}, \mu_2 +\delta_{k, 1} + \delta_{k,3},t\right) \, .
\label{matF}
\end{equation}
As we are interested in evaluating the properties of the two-modes of the electromagnetic field in the cavity, it is convenient to take the trace of the density matrix of $\rho^{(F)}_{\nu_1,\nu_2; \mu_1,\mu_2} (t)$ with respect to the first and second modes of the field, that is, 
\begin{equation}
\rho^{(F_1)}_{\nu_1,\mu_1} =\sum_{\nu_2} \rho^{(F)}_{\nu_1,\nu_2; \mu_1,\nu_2} (t) \, , \qquad \rho^{(F_2)}_{\nu_2,\mu_2} =\sum_{\nu_1} \rho^{(F)}_{\nu_1,\nu_2; \nu_1,\nu_2} (t) \, .
\label{redf1f2}
\end{equation}
The last expression allows us to study the behavior of the modes of the field in phase space.

Therefore we are going to calculate  the Husimi and/or the Wigner quasiprobability distribution functions. The Husimi function $\mathcal{Q}_H$ is the expectation value of the reduced densities $\hat{\rho}^{(F_k)}$ with respect to the coherent basis states $\ket{\beta}$, 
\begin{equation}
\mathcal{Q}_{H_k}=\frac{1}{\pi}\braket{\beta|\hat{\rho}^{(F_k)}|\beta} \, ,
\end{equation} 
with $k=1,2$, and the complex parameter beta was defined in the last section in terms of the quadrature amplitudes. Following its definition for both field modes, we get,
\begin{eqnarray}
\mathcal{Q}_{H_k}&=&\frac{\mathrm{e}^{-\left(X^{2}+Y^{2}\right)}}{\pi}\sum_{\substack{\nu_1\, ,  \mu_1}} \frac{\left(X- i\,  Y\right)^{\nu_{1}}\left(X+ i \, Y\right)^{\mu_{1}}}{\sqrt{\nu_{1}! \, \mu_{1}!}} \hat{\rho}^{(F_{k)}}_{\nu_1,\mu_1} \, .
\end{eqnarray}

The second order coherence is useful to characterize the different kinds of light, and an alternative measure is provided by the Mandel parameter $Q_M$, defined as
\begin{equation}
\mathcal{Q}_{M_k}=\frac{\braket{\hat{n}_k^2}-\braket{\hat{n}_k}^2}{\braket{\hat{n}_k}}-1 \,, \label{Mandel}
\end{equation} 
with $k=1,2$. For each mode, we can study how light-atom interaction modifies the field distribution. A super-Poissonian statics is characterized by a variance $\Delta n_k^2 > \braket{\hat{n}_k}$. Commonly, this distribution is observed in classical electromagnetic fields, such as thermal light with a geometric distribution of photons.   Sub-Poisson statistics is defined by $\Delta \hat{n}_k^2 < \braket{\hat{n}_k}$ and it is associated with quantized light (Fock states of light), i.e., it cannot be described by the classical electromagnetic theory. Finally $Q_{M_k}=1$ denotes coherent light, characterized by a Poissonian distribution.

\subsubsection{Dark states}

Finally, we resume the discussion of the set of dark states \eqref{eq:darkstates}. Despite wavefunction $\psi(t)$ was constructed on a basis state which does not explicitly consider the dark states, the probability of the system of being in the dark states with a fixed number of photons $n$ is different from zero and is given by  
\begin{eqnarray}
{\cal P}_{d_1}(n)&=&\left|\langle\psi_{d_1}(n) | \psi(t)\right|^2= \left|G_1(\alpha_s,\hat{\gamma}_1,n, n+1,t)\right|^2 = e^{-|\alpha_1|^2} \, |\hat{\gamma}_1|^2 \, e^{-|\alpha_2|^2} \, \frac{|\alpha_2|^{2\, n}}{n!} \, , \nonumber \\
 {\cal P}_{d_2}(n)&=&\left|\langle \psi_{d_2}(n) | \psi(t)\right|^2=\left |G_2(\alpha_s,\hat{\gamma}_2, n,0,t)\right|^2= e^{-|\alpha_2|^2} \, |\hat{\gamma}_2|^2 \, e^{-|\alpha_1|^2} \, \frac{|\alpha_1|^{2\, n}}{n!} \,  .
 \label{eqdark}
\end{eqnarray}
Comparing the last expressions with the joint probability of having $m_1$ and $m_2$ excitations for the corresponding constants of the motion~(see, Eq.~(\ref{probM1M2})), it is concluded that ${\cal P}_{d_1}(n)={\cal P}(m,m+1,t)$ and ${\cal P}_{d_2}(n)={\cal P}(m,0,t)$. 

Notice that both expressions in~(\ref{eqdark}) are multiplied by a Poissonian photon distribution function for one or the other modes of the electromagnetic field in the cavity. Then it is straightforward that the probability of the system to be in the dark states \eqref{eq:darkstates} independently of the number of photons is given by,
\begin{equation}
{\cal P}_{d_1} = \sum_n \, {\cal P}_{d_1}(n)=e^{-|\alpha_1|^2} \, |\hat{\gamma}_1|^2 \, \, , \quad
{\cal P}_{d_2}=\sum_n \, {\cal P}_{d_2}(n)= e^{-|\alpha_2|^2} \, |\hat{\gamma}_2|^2 \,  \,  .
\end{equation}
The eigenvalues of the constants of the motion $\hat{M}_1$ and $\hat{M}_2$ for the dark states are the following: for the state $|\psi_{d_1}(n)\rangle$ one has $m_1=n$ and $m_2=n+1$ while for $|\psi_{d_2}(n)\rangle$ $m_1=n$ and $m_2=0$, as it was mentioned above.


\section{Results}

\begin{figure}
\begin{centering}
\includegraphics[scale=0.6]{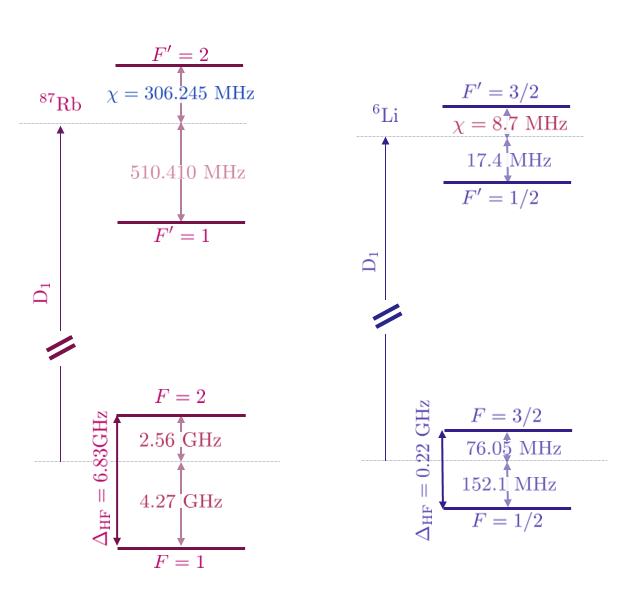}
\end{centering}
\caption{\label{fig:3:D1RbLi} Hyperfine structure for the isotopes $^6$Li and $^{87}$Rb.}
\end{figure}

The Hamiltonian model \eqref{eq:Paper-001} can be applied for the experimental data of the $D_1$ transition in alkaline atoms, in spite of that they exhibit a very different separation between their energy levels. We show how the model is applied and that their behavior is essentially the same. To illustrate this, we solve the model for the isotopes $^6\mathrm{Li}$ and $^{87}\mathrm{Rb}$ (Figure \ref{fig:3:D1RbLi}). In Table~\ref{table:parametersLiRb}, the experimental data for the $D_1$ transition are established, that is, the natural linewidth $\Gamma$, the wavelength $\lambda^{D_1}$ and the frequency $\nu^{D_1}$ of the $D_1$ line, and the saturation intensity $I^{D_1}_{\rm sat}$. We are going to consider the $D_1$ transition line although a similar procedure can be used for the $D_2$ transition line. 

\begin{table}[h]
    \caption{Atomic Data for the $D_1$ transition line in $^6$Li and $^{87}$Rb}
    \begin{subtable}{.5\linewidth}
      \centering
        \subcaption{ $^6$Li Data}
        \begin{tabular}{||c c c ||}
            \hline
 \textbf{Property} & \textbf{Symbol} & \textbf{Value}  \\ [0.5ex] 
 \hline\hline
 Natural Linewidth & $\Gamma$(Li) & $2 \pi\, 5.8724$ MHz \\ 
 \hline
 $\mathrm{D_1}$ wavelength & $\lambda^{D_1}(\mathrm{Li})$ & 670.992421 nm \\
 \hline
 $\mathrm{D_1}$ frequency & $\nu^{D_1}(\mathrm{Li})$ & 446.789634 THz\\
 \hline
 $\mathrm{D_1}$ saturation intensity & $I_{sat}^{D_1}(\mathrm{Li})$ &  $7.59\,\mathrm{mW/cm^2}$ \\
 [1ex] 
 \hline
        \end{tabular} 
    \end{subtable}%
    \begin{subtable}{.5\linewidth}
      \centering
        \subcaption{$^{87}$Rb Data}
        \begin{tabular}{||c c c ||}
          \hline
 \textbf{Property} & \textbf{Symbol} & \textbf{Value}  \\ [0.5ex] 
 \hline\hline
 Natural Linewidth & $\Gamma$(Rb) & $2 \pi\, 5.746$ MHz \\ 
 \hline
 $\mathrm{D_1}$ wavelength & $\lambda^{D_1}(\mathrm{Rb})$ & 794.978850 nm \\
 \hline
 $\mathrm{D_1}$ frequency & $\nu^{D_1}(\mathrm{Rb})$ & 337.1074635 THz\\
 \hline
 $\mathrm{D_1}$ saturation intensity & $I_{sat}^{D_1}(\mathrm{Rb})$ &  $4.484(5)\,\mathrm{mW/cm^2}$ \\
 [1ex] 
 \hline
        \end{tabular}
    \end{subtable} 
    \label{table:parametersLiRb}
\end{table}

Once a fine structure transition ($D_1$ or $D_2$) is chosen, generating the appropiate frequencies to detect the hyperfine levels is necessary. As illustrated in Figure \ref{fig:3:D1RbLi}, the hyperfine separation of the ground levels in $^6\mathrm{Li}$ is of the order of MHz, while for $^{87}\mathrm{Rb}$, the splitting is of the order of GHz. This difference in the frequency or energy will be reflected in the delay or advance of the emergency of the collapse and revival phenomena, as it will be discussed later. Experimentally the separation in $^6\mathrm{Li}$ is achieved by  optical devices \cite{Weiner2003, McAlexander1996, Scherf1996, Windholz1992}. For $^{87}\mathrm{Rb}$, two lasers with a difference in frequency of 6.8 GHZ must be used \cite{Weiner2003, Sagle1996, Siddons2008}.  
 
By a appropiate choice of frequencies, the hyperfine structure of the alkaline atom can be simplified to the three-level system model (Figure \ref{fig:3:level_model}). Due to the very small energetic difference of the hyperfine levels of the $^2P_{1/2}$ state of $^6\mathrm{Li}$, there is no experimental resolution for the $F$ levels. Thus, an effective excited state can be considered. In $^{87}\mathrm{Rb}$, the separation of the F levels of the excited state $^2P_{1/2}$ are bigger and are accessible experimentally. We chose a transition to the excited state $F=2$.  With these considerations, the atomic frequencies are given by:  
\begin{eqnarray}
\omega_1(\mathrm{Li})\!&=& \! 2\pi \left[\nu^{D_1}(\mathrm{Li}) + 152.1 \,\mathrm{MHz}\right],\quad \omega_2(\mathrm{Li})=2\pi \left[ \nu^{D_1}(\mathrm{Li}) - \tfrac{152.1}{2} \,\mathrm{MHz}\right],\quad \omega_3(\mathrm{Li})= 2\pi  \nu^{D_1}(\mathrm{Li}) \nonumber \\
\omega_1(\mathrm{Rb})\! &=&\! 2\pi \left[ \nu^{D_1}(\mathrm{Rb})+\chi + 4.2717\, \mathrm{GHz}\right],\quad \omega_2(\mathrm{Rb})=2\pi \left[ \nu^{D_1}(\mathrm{Rb})+\chi - 5.563\, \mathrm{GHz} \right],\quad \omega_3(\mathrm{Rb})=2\pi \left[\nu^{D_1}(\mathrm{Rb})+\chi \right]\, ,\nonumber
\end{eqnarray}
\emph{•}where $\chi=306.24$ MHz is the separation of the excited state with $F=2$. As discussed before, fine transition $D_1$ is the reference for the atoms, but the dynamic and the behavior of the atom-field interaction is a consequence of the hyperfine states separations. Thus, it is possible to make a global shift in the frequencies by considering just these small shifts of MHz and GHz for $^6\mathrm{Li}$ and $^{87}\mathrm{Rb}$, respectively.  By taking out the values of the fine states, the shifted frequencies have the value:
\begin{eqnarray}
\omega_1^{\prime}(\mathrm{Li})\!&=& \! 2\pi \cdot 152.1 \quad \mathrm{MHz},\quad \omega_2^{\prime}(\mathrm{Li})= -2\pi \cdot \tfrac{152.1}{2} \ \, \mathrm{MHz},\quad \omega_3^{\prime}(\mathrm{Li})= 0 \nonumber \\
\omega_1^{\prime}(\mathrm{Rb})\! &=&\! 2\pi \cdot 4.2717\, \mathrm{GHz},\quad \omega_2^{\prime}(\mathrm{Rb})=-2\pi \cdot 2.563\, \mathrm{GHz},\quad \omega_3^{\prime}(\mathrm{Rb})=0 ,\nonumber
\end{eqnarray}
Next, we added the atomic frequencies $2\pi \times \tfrac{152.1}{2}$ MHz for lithium and $2 \pi \times$ 5.563 GHz for rubidium to remove the negative values and rearranged the labels to satisfy the order of our model: $ \omega_1 < \omega_2 < \omega_3$. Table \ref{table:adimensional_frequencies} shows the final frequencies of the system and its adimensional values, which were applied in our results. Notice that the frequencies are measured in the corresponding units of $\omega_3$.

For an atom coupled to a near-resonant optical field, the frequency at which the interaction coherently drives the atom between two of its states is known as Rabi frequency. In a semiclassical approximation, this is defined as: 
\begin{equation}
\mu=\frac{\braket{b|\mathbf{d}\cdot \mathbf{E}|a}}{h}=\frac{d_{ba}E_0}{h}\,,
\end{equation} 
where $d_{ba}$ is the electric-dipole transition matrix element for states $a$ and $b$, and $E_0$ is the electric field strenght of the inicident optical field. We want to express the Rabi frequency in terms of experimental parameters that can be modulated. Thus, it is useful to introduce the definition of saturation intensity, which is defined as the intensity at which a monochromatic beam excites the transition at a rate equal to one-half of its natural line width: 
$$ I_{sat}=\frac{c \epsilon_0 \Gamma^2 \hbar^2}{4 |\mathbf{d} \cdot \mathbf{e} |^2 },$$ 
where $\mathbf{e}$ is the unit polarization vector, $\mathbf{E}=\mathbf{e}E_0$. Thus, we can establish the relation: $\frac{I}{I_{sat}}=2\left( \frac{\mu}{\Gamma}\right)^2$.
The Rabi frequencies for the two allowed transitions in the system are: 
\begin{equation}
\mu_{13}=\Gamma \sqrt{\frac{I_1}{2 I_{sat}}}\, , \qquad \mu_{23}=\Gamma \sqrt{\frac{I_2}{2 I_{sat}}},
\end{equation}
where $I_1$ and $I_2$ are the intensities of the fields for transitions $\ket{1_A}\leftrightarrow \ket{3_A}$ and $\ket{2_A}\leftrightarrow \ket{3_A}$, respectively.
 $\Gamma$ is the spontaneous decay rate of the transition and $I_{sat}$ is known as the intensity saturation, defined as the magnitude of $I$ needed to make the Rabi frequency equal to $\Gamma$. The relation between these two last values is important because when $\mu$ is less than the spontaneous decay rate, the atom is likely to spontaneously decay out of the excited state. As the intensity of the applied field increases, the atom-light interaction becomes stronger, and the atom is driven coherently by the applied field. Another parameter that can be switched experimentally is detuning, $ \Delta_{13}=\Delta_{23}$, which we expressed in terms of $\Gamma$:
 \begin{equation*}
 \Delta_{13}=n\Gamma,
 \end{equation*}
where $n$ is a real number that can take negative or positive values. The closer $n$ is to zero, the atom-field interaction is said to be resonant, and the principal contribution is a radiation force in which an interchange of momentum between atom and photons exists. While $n$ is bigger, the atomic transitions are less able to occur. 

\begin{table}
\caption{Frequencies for the three level system of Li and Rb are expressed in units of the corresponding $\omega_3$. Therefore the unit of time for Li is $0.698$ ns while for Rb one has $2.093$ ns.
\label{table:adimensional_frequencies}}
\begin{tabular}{||c c || c c ||}
 \hline
 \textbf{Frequency} & \textbf{Value} &\textbf{Adimensional Frequency} & \textbf{Value} \\ [0.5ex] 
 \hline\hline
$\Gamma$(Li) & $2\pi$ 5.8724 MHz & $\bar{\Gamma}$(Li)  & 0.0257 \\ 
 \hline
$\omega_1$(Li)  &  0  &  $\bar{\omega}_1$(Li)  & 0\\ 
 \hline
$\omega_2$(Li) & $ 2\pi \frac{152.1}{2}$ MHz &  $\bar{\omega}_2$(Li) & $\frac{1}{3}$\\
 \hline
$\omega_3$(Li) &   $2\pi\frac{3\times 152.1}{2}$  MHz   &   $\bar{\omega}_3$(Li) & 1.0\\
 \hline
 \hline
$\Gamma$(Rb) &  $2\pi$ 5.746 MHz  &   $\bar{\Gamma}$(Rb)  & $8.407\times 10^{-4}$ \\ 
  \hline
$\omega_1$(Rb) & 0 &  $\bar{\omega}_1$(Rb) &  0 \\
\hline
$\omega_2$(Rb)& $2\pi$ 2.563 GHz &  $\bar{\omega}_2$(Rb) &  0.375 \\
\hline
$\omega_3$(Rb)  & $2\pi$ 6.835 GHz &  $\bar{\omega}_3$(Rb) &  1.0 \\
 [1ex] 
 \hline
\end{tabular}
\end{table}

\subsection{Dynamic properties}

We study the temporal evolution of the system by considering initial states build of the tensorial product $\ket{\psi_0}=\ket{\alpha_1,\alpha_2}\times\ket{\phi_M}$, where $|\alpha_k|^2$ is determined by the average photon number of mode $k$ in the cavity. We consider for the matter sector the set of four states,
\begin{equation}
\{ \ket{\phi_M} \} =\{ \ket{1_A},\quad \ket{2_A},\quad \ket{3_A},\quad \frac{1}{\sqrt{3}}\left(\ket{1_A}+\ket{2_A}+\ket{3_A}\right) \}\,. \label{initial_states}
\end{equation}
We fix the mean value of photons in the cavity to be $\braket{\hat{n}_1}=\braket{\hat{n}_2}=3$ and the intensities of the beams of light with the values $I_1=I_2= 3 \, I_{sat}$. These intensities also define the Rabi frequencies $\mu_{13}=\mu_{23}=\Gamma \sqrt{\frac{3}{2}}$. The zero detuning, $\Delta_{13}=0$, is considered so that the atom is in resonance with both modes of the field, except for the initial state with $\ket{\phi_M}=\ket{2_A}$, where the case $\Delta_{13}=5\,\Gamma$ is considered. 

Additionally, we study two sets of parameters that yield a suppression of one of its atomic transitions. One of these cases has been considered by Gerry and Eberly, that is, the excited state is considered to be far off resonance and then it is adiabatically eliminated \cite{Gerry1990}.  For this reason, we call them \textit{Raman-I}  when we take the set of parameters 
$$\{I_1=5 \, I_{sat},I_2=\tfrac{1}{4} \, I_{sat},\braket{\hat{n}_1}=3,\braket{\hat{n}_2}=1,\Delta_{13}=0,\ket{\psi_0}=\ket{\alpha_1,\alpha_2}\times \ket{3_A}\} \, , $$
and \textit{Raman-II} when the parameters are 
$$\{ I_1=I_2=I_{sat},\braket{\hat{n}_1}=\braket{\hat{n}_2}=3,\Delta_{13}=10\, \Gamma,\ket{\psi_0}=\ket{\alpha_1,\alpha_2}\times \ket{1_A} \} \, . $$

\begin{figure}[ht]
\centering
\begin{subfigure}{0.47\textwidth}
    \includegraphics[width=\textwidth]{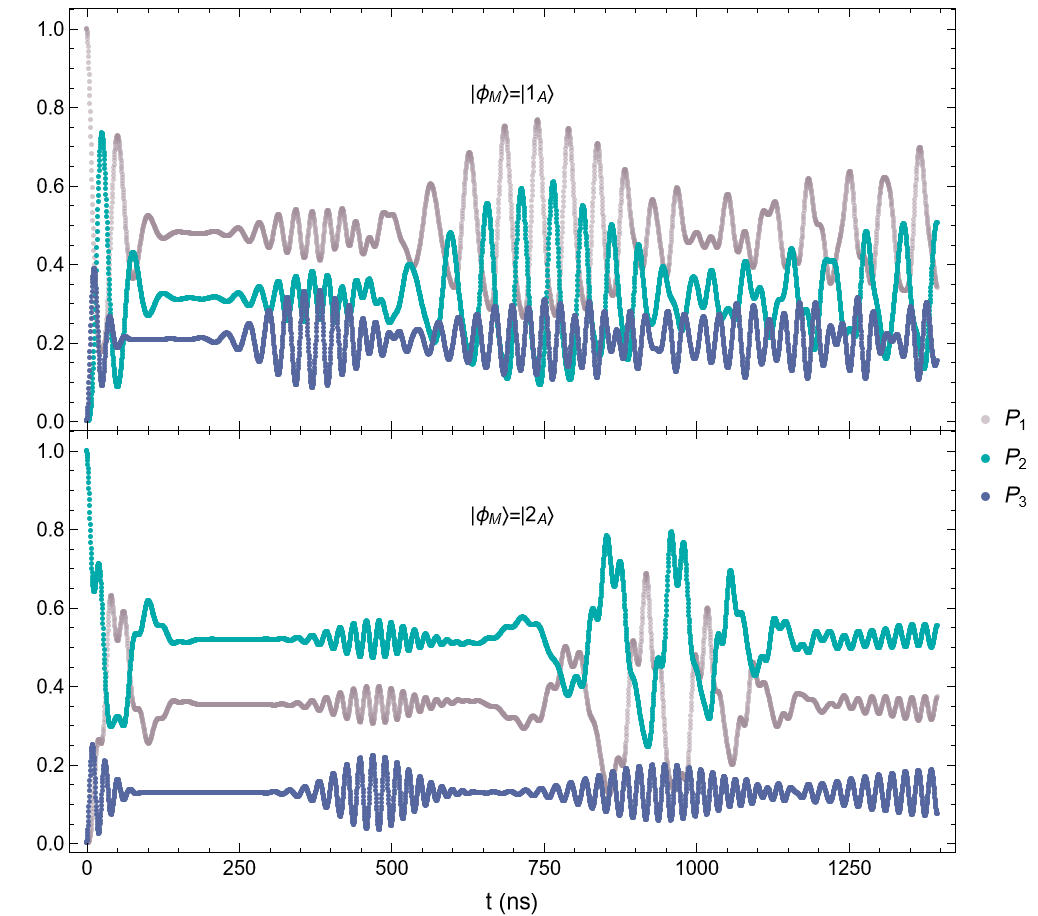}
 \caption{ }
    \label{fig:first}
\end{subfigure}
\hfill
\begin{subfigure}{0.47\textwidth}
    \includegraphics[width=\textwidth]{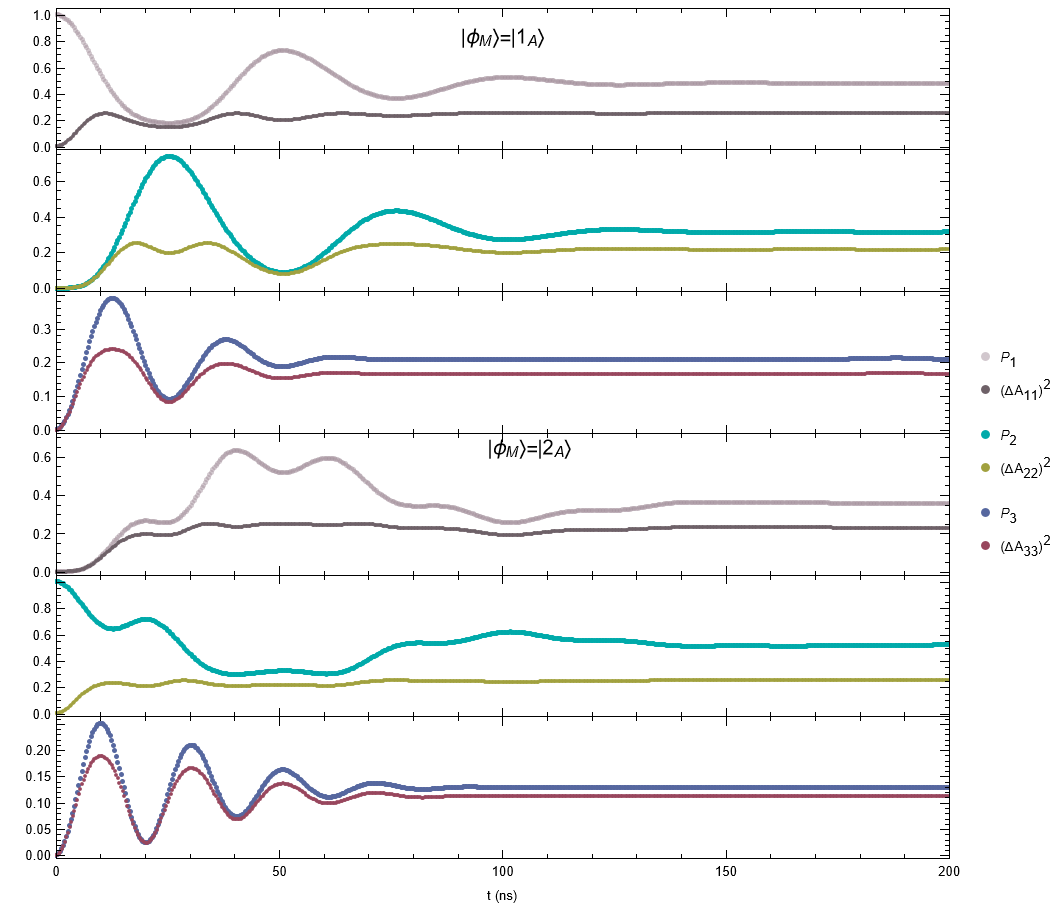}
    \caption{}
    \label{fig:second}
\end{subfigure}
\hfill        
\begin{subfigure}{0.47\textwidth}
    \includegraphics[width=\textwidth]{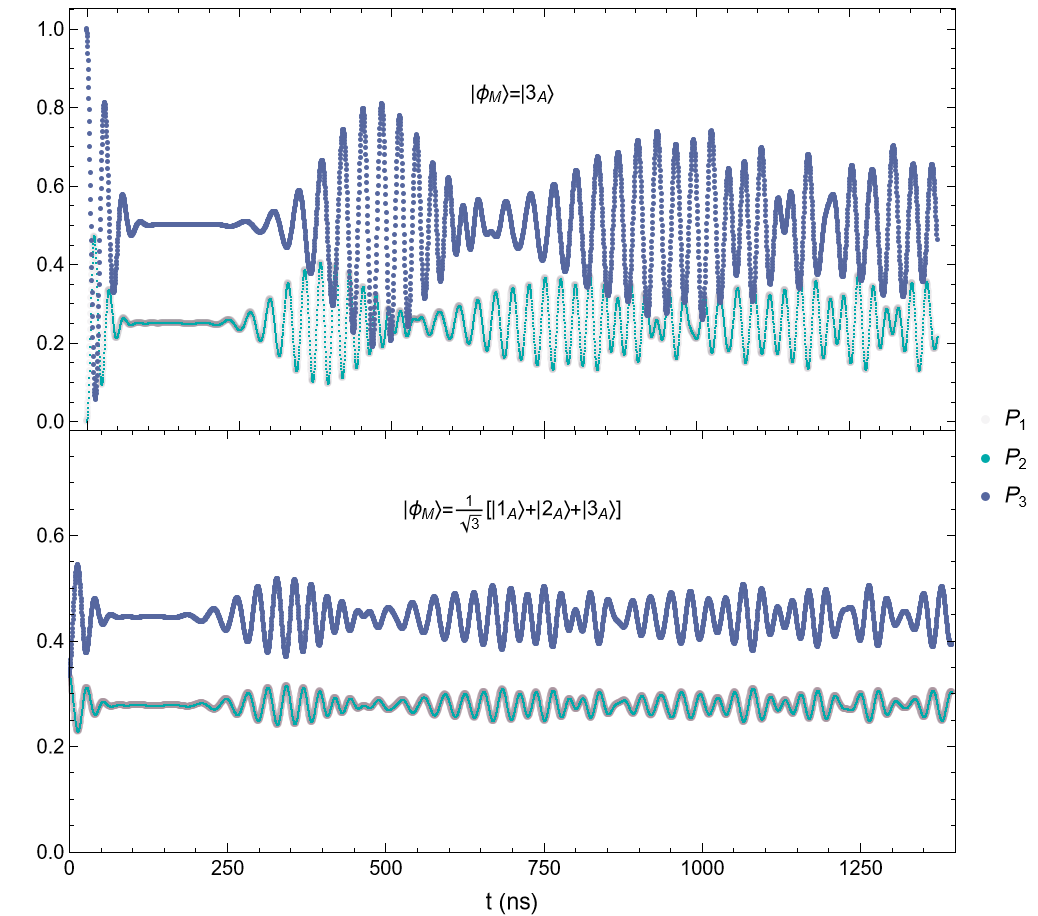}
    \caption{}
    \label{fig:third}
\end{subfigure}    
\hfill
\begin{subfigure}{0.47\textwidth}
    \includegraphics[width=\textwidth]{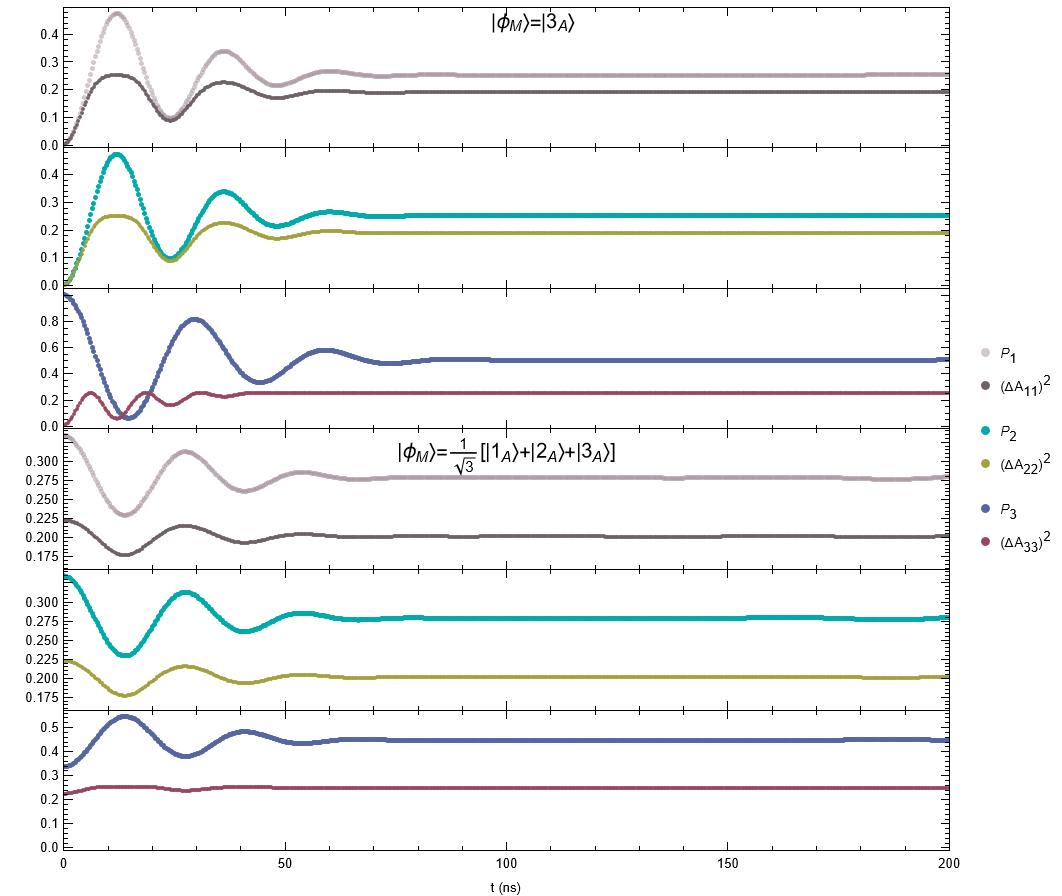}
    \caption{}
    \label{fig:fourth}
\end{subfigure}  
\label{fig:ProbAtom1}
\caption{Evolution of the atomic occupation probabilities $\mathcal{P}_k$ and its corresponding fluctuations $(\Delta\braket{\hat{A}_{kk}})^2$, with $k=1,2,3$, are shown as a function of time for initial states of the form $\ket{\psi_0}=\ket{\alpha_1,\alpha_2}\times \ket{\phi_M}$. In (a) for the initial atomic states $\ket{\phi_M}=\ket{1_A}$ and $\ket{\phi_M}=\ket{2_A}$ and in (b) the corresponding fluctuations. In (c) the atomic states are $\ket{\phi_M}=\ket{3_A}$ and $\ket{\phi_M}=\tfrac{1}{\sqrt{3}}[\ket{1_A}+\ket{2_A}+\ket{3_A}]$ and (d) shows the associated fluctuations. The parameters used are the following: $I_1=I_2=3I_{sat}$, $\braket{\hat{n}_1}=\braket{\hat{n}_2}=3$, and $\Delta_{13}=0$ except for the initial state $\ket{2_A}$ where use $\Delta_{13}=5\Gamma$.  }
\end{figure}

In figures~\ref{fig:first}~to~\ref{fig:fourth}, the evolution of the atomic occupation probabilities $\mathcal{P}_k$ and its corresponding fluctuations $(\Delta\braket{\hat{A}_{kk}})^2$ are shown, with $k=1,2,3$.  The considered initial states are given in expression~\eqref{initial_states} together with two-mode coherent states for the electromagnetic fied in the cavity. 
 
In Figure~\ref{fig:first} the probabilities are displayed for the states $\ket{\phi_M}=\ket{1_A}$ and $\ket{\phi_M}=\ket{2_A}$. For the case $\ket{1_A}$, we consider zero detuning, $\Delta_{13}=0$ while for $\ket{2_A}$, we take $\Delta_{13}=5 \, \Gamma$. For the same resonant condition, the evolution of $\mathcal{P}_1$ and $\mathcal{P}_2$ are only interchanged and have the same behavior, together with $\mathcal{P}_3$. Due to the change in the field frequency for the non-resonant case, the coupling of the two subsystems is less than in the resonant condition. This can be observed in the collapse and revivals patterns, which present a decrease in their oscillations. Figure \ref{fig:second} shows the corresponding fluctuations of the atomic populations in Figure \ref{fig:first}. Notice that, occupation probabilities and fluctuations values are of the same order of magnitude, and even, at certain times, they are equal. Due to the definition of $(\Delta\braket{\hat{A}_{kk}})^2\equiv(\Delta\mathcal{P}_k)^2=\mathcal{P}_k(1-\mathcal{P}_k)$, which arises because of the idempotent property of operator $\hat{A}_{kk}$,  we see that probabilities and fluctuations must be very similar when $\mathcal{P}_k$ tends to zero. 

Figure \ref{fig:third} shows the atomic occupation probabilites for the initial atomic states $\ket{3_A}$ and $\tfrac{1}{\sqrt{3}}\left(\ket{1_A}+\ket{2_A}+\ket{3_A}\right)$.  If we compare them with the other two initial states, we observe that $\mathcal{P}_1$ and $\mathcal{P}_2$ behave the same in spite of the difference in the energy of the atomic levels $\hbar \omega_1$ and $\hbar \omega_2$. Thus, for these initial conditions, atomic states $\ket{1_A}$ and $\ket{2_A}$ have the same stability and are equally populated. Figure \ref{fig:fourth} shows the associated fluctuations, as expected, when the initial atomic state is given by a linear combination of basis states, at $t=0$ the fluctuations are different from zero.

\begin{figure}
\centering
\begin{subfigure}{0.47\textwidth}
    \includegraphics[width=\textwidth]{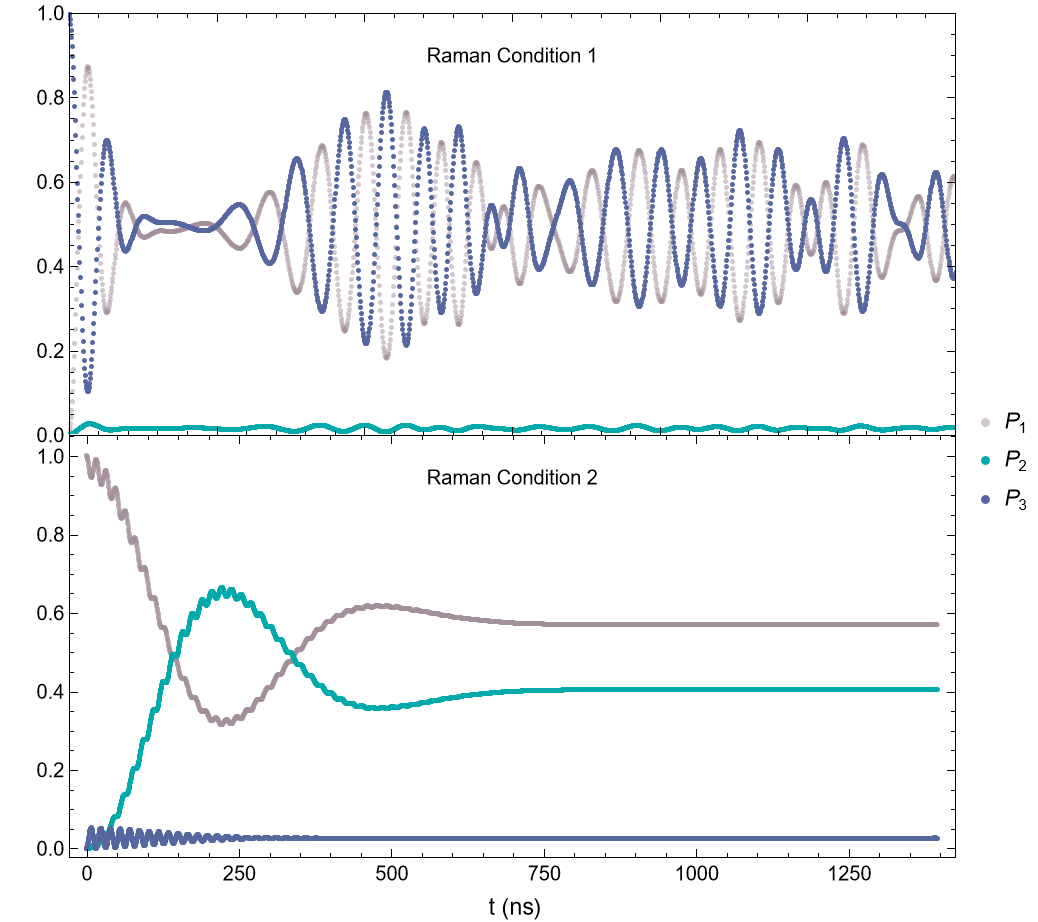}
\end{subfigure}
\hfill
\begin{subfigure}{0.47\textwidth}
    \includegraphics[width=\textwidth]{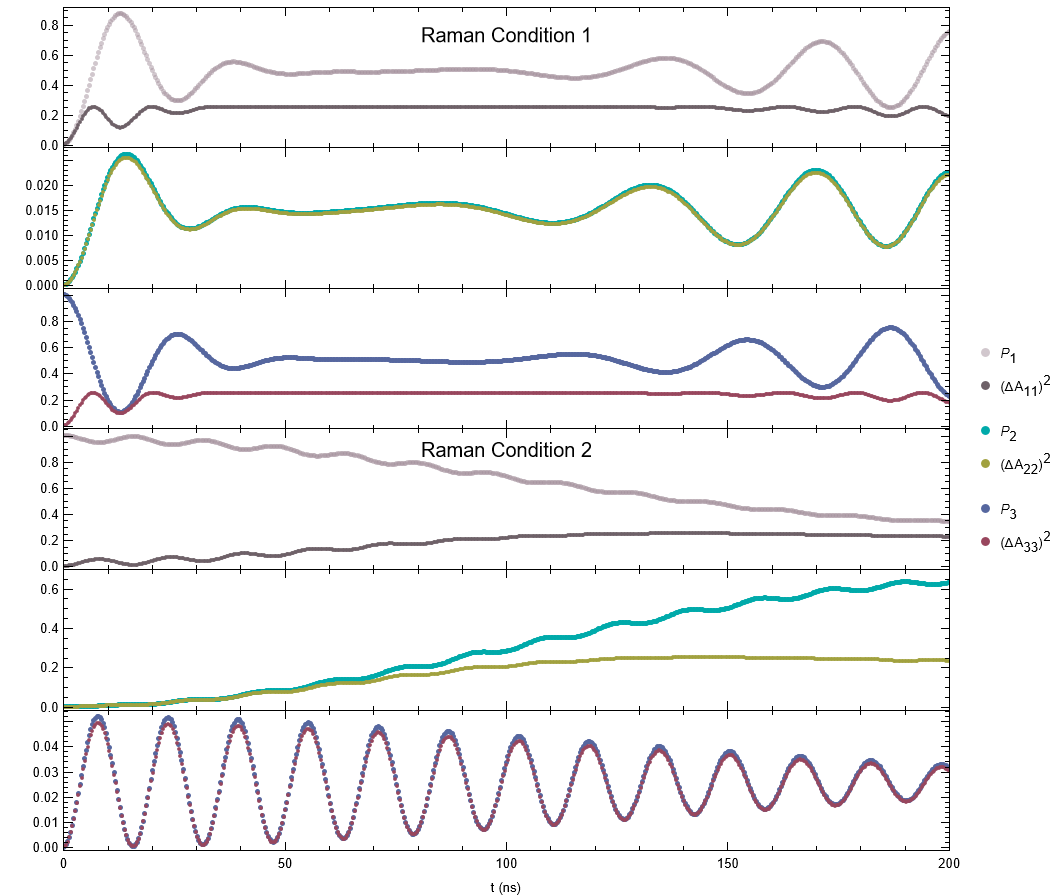}
\end{subfigure}
\caption{The temporal evolution of the atomic occupation probabilities $\mathcal{P}_k$ and its corresponding fluctuations $(\Delta\braket{\hat{A}_{kk}})^2$ are shown, with $k=1,2,3$. At top we use the parameters of the {\it Raman-I} and at bottom those of \textit{Raman-II} \hfill}
\label{fig:Raman1}
\end{figure}

The occupation probabilities for the Raman Conditions and the associated fluctuations are displayed in Figure \ref{fig:Raman1}. For the case \textit{Raman-I}, the population $\mathcal{P}_2$ goes to zero implying a Raman transition, however the transition $\ket{2_A}\rightarrow \ket{3_A}$ is suppressed due to the weak coupling between the atom with the mode 2 of the electromagnetic field.  For \textit{Raman-II}, a Raman transition exists in which the probability of occupation for state $\ket{3_A}$ is very small compared with the occupation probabilities of the states $\ket{1_A}$ and $\ket{2_A}$ and then one can say that the third level behaves as a virtual level. This behavior is achieved with a very large detuning, that is, with $\Delta_{13}^2>\mu_{13}^2$ or $\Delta_{13}^2>\mu_{23}^2 $. We consider $\Delta_{13} \geq 10\sqrt{2} \, \mu_{ij}$ in order to reach this condition, with $\mu_{ij}=\mu_{13}$ or $\mu{ij}=\mu_{23}$ we get the same result.

\begin{figure}[ht]
\begin{subfigure}{0.47\textwidth}
    \includegraphics[width=\textwidth]{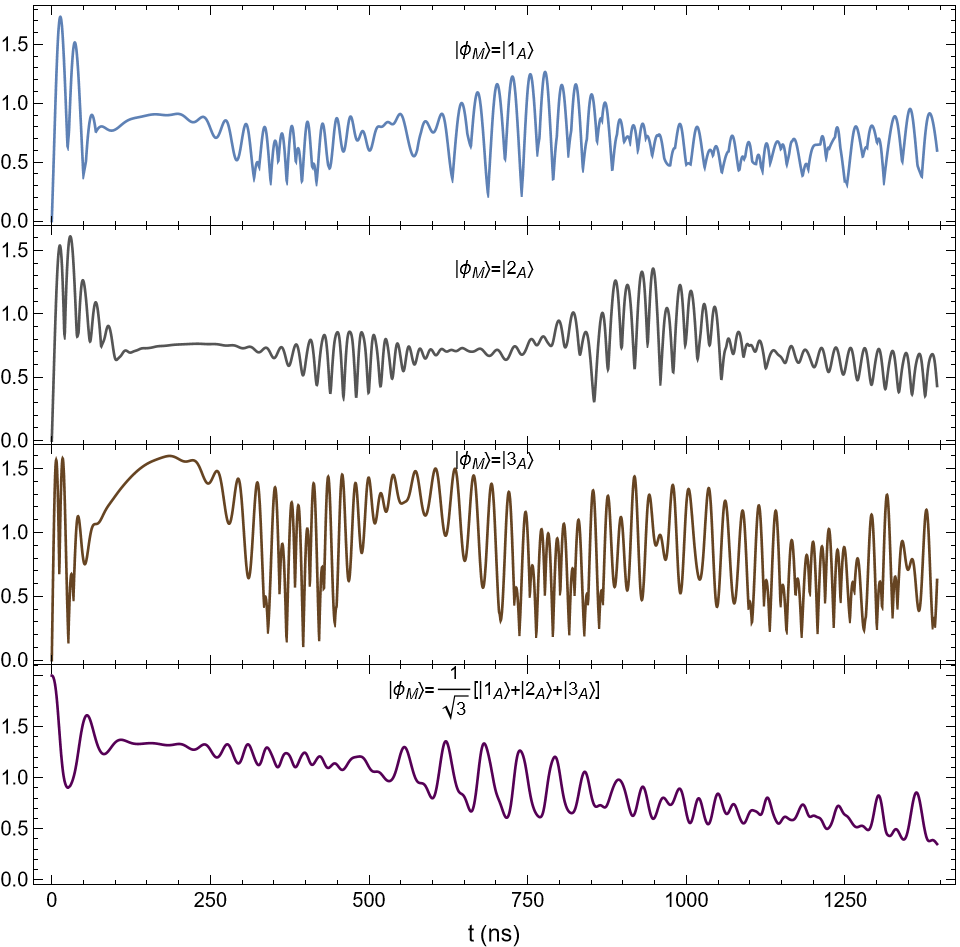}
\caption{}
\label{fig:Coh_a}
\end{subfigure}
\hfill
\begin{subfigure}{0.47\textwidth}
    \includegraphics[width=\textwidth]{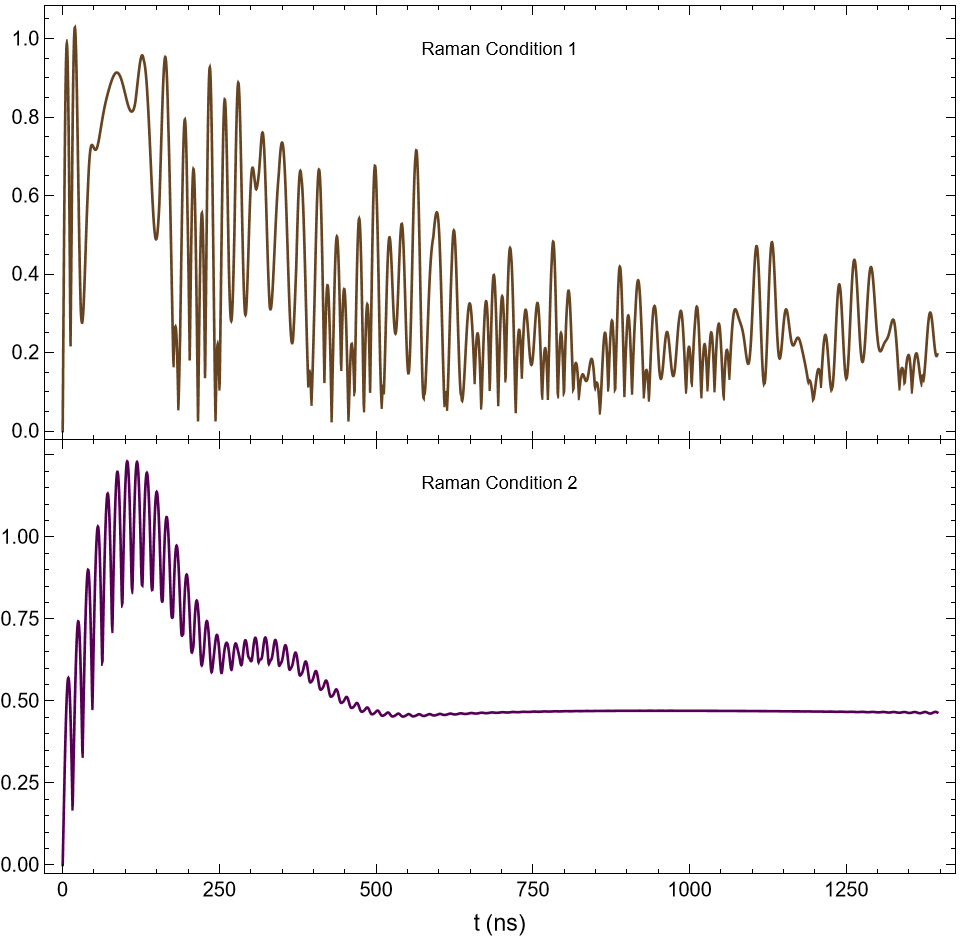}
\caption{}
\label{fig:Coh_b}
\end{subfigure}
\caption{ Time evolution of coherence function \eqref{CoherenceFunction}. In a)  
the behavior of the coherence for the four initial atomic states with the fixed field conditions considered for these cases. In b) the behavior of coherence when Raman Conditions are taken into account \hfill  
}
\label{fig:Coh}
\end{figure}

The coherence function given in expression~\eqref{CoherenceFunction} is associated to the reduced density matrix of the matter and establishes the correlation or transitions between the energy levels of the system. Their temporal evolution is exhibited in Figure \ref{fig:Coh_a}, for different initial conditions and using the parameters of the model mentioned above. One sees at the same times where the collapses appear, that the coherence function is not an oscillating function. In Figure \ref{fig:Coh_b} the coherence for the Raman conditions as functions of time are displayed.  In all the cases, the coherence function never goes to zero, except at the initial time for the atomic states $|1_A \rangle$, $|2_A\rangle$ and $|3_A\rangle$. Notice that, the case \textit{Raman-II} stabilizes the coherence function in $1/2$. In summary, the oscillations of the functions match the collapses and revivals patterns observed in the atomic occupation probabilities. 

\begin{figure}[h]
\begin{subfigure}{0.47\textwidth}
    \includegraphics[width=\textwidth]{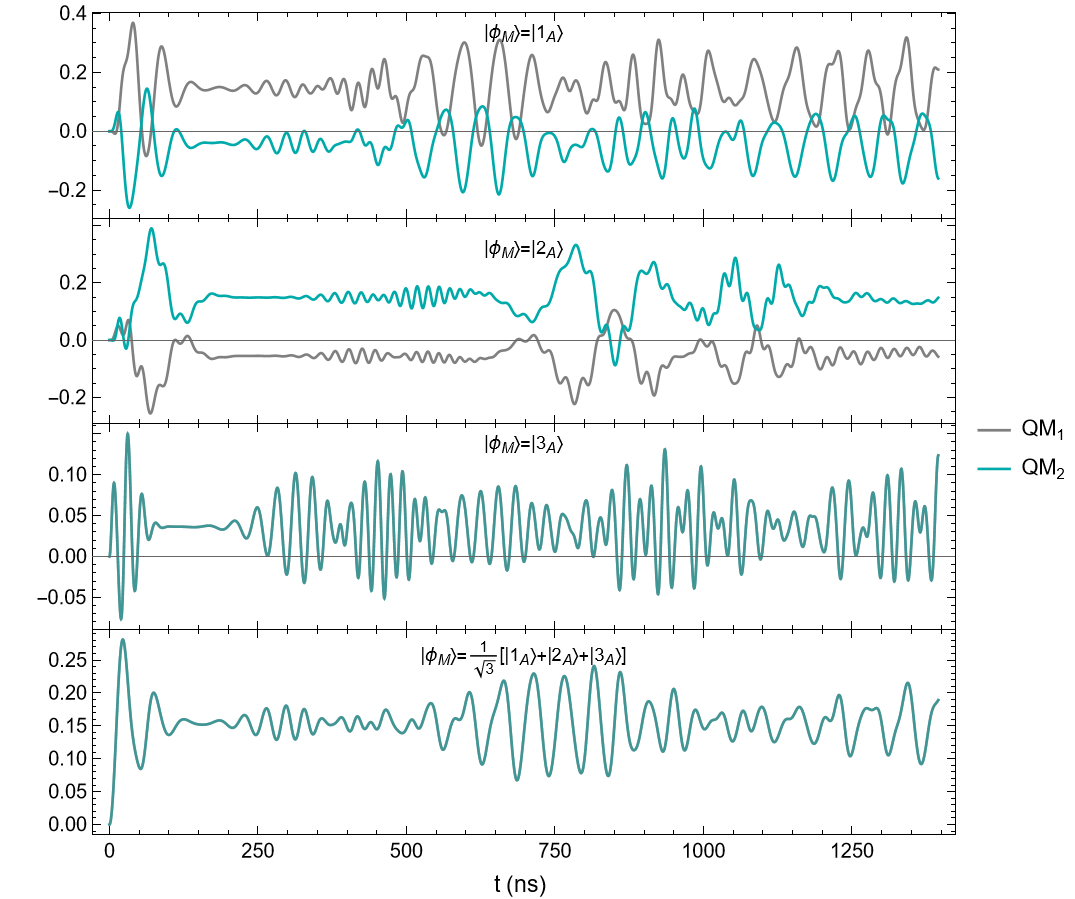}
   \caption{}
    \label{fig:QM_a}
\end{subfigure}
\hfill
\begin{subfigure}{0.49\textwidth}
    \includegraphics[width=\textwidth]{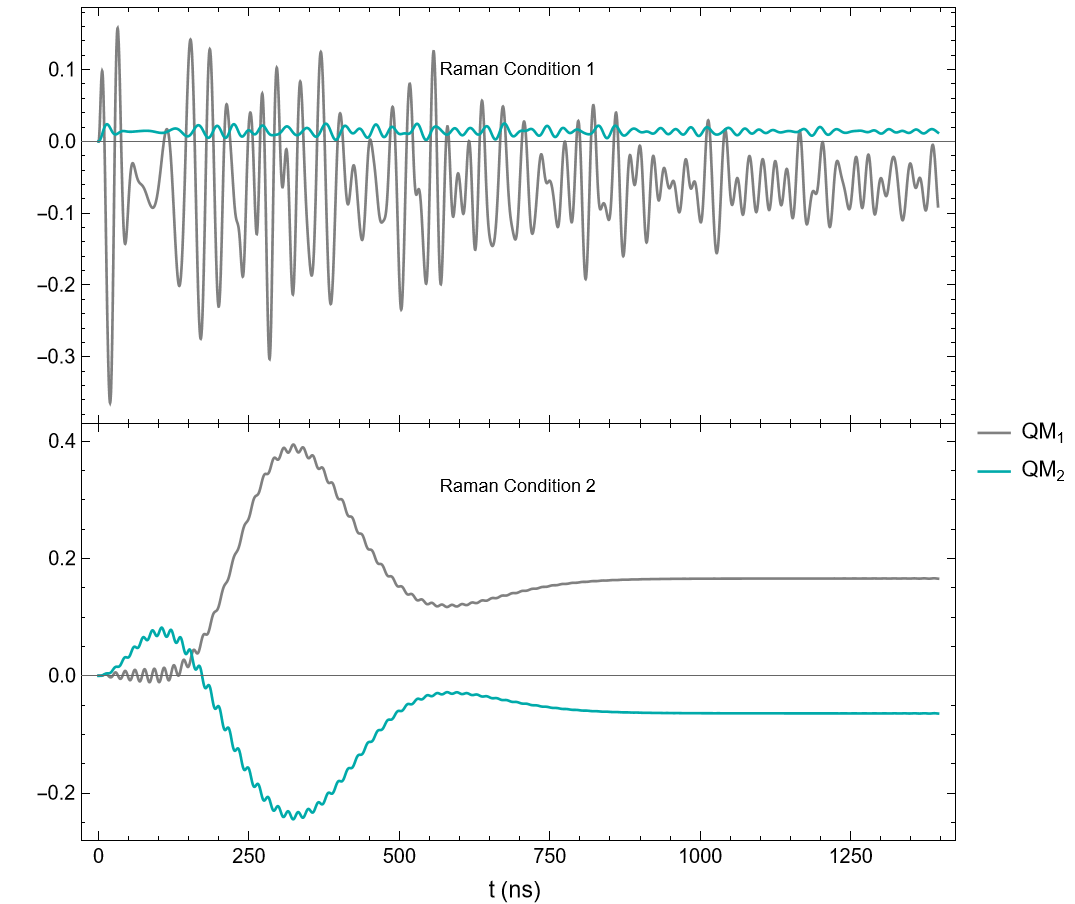}
    \caption{}
    \label{fig:QM_b}
\end{subfigure}
 \caption{ Evolution of the Mandel parameters \eqref{Mandel} for photons with frequency $\Omega_1$, $QM_1(t)$, and for photons with frequency $\Omega_2$, $QM_2(t)$, as a function of time. In a) the behavior when the initial states \eqref{initial_states} and fields with an average number of photons $\bar{n}_1=\bar{n}_2=3$. In b) the evolution of the Mandel parameters when Raman conditions are taken in account. \hfill }
\label{fig:QM}
\end{figure}

Atom-field interaction modifies the statistical properties of the photon modes of the electromagnetic field in the cavity. The Mandel parameter \eqref{Mandel} allows us to measure these changes and additionally determine their photon distribution function.  The evolution of the Madel parameter is indicated in Figure~\ref{fig:QM}, most of the time for the atomic states $|1_A\rangle$ and $|2_A\rangle$ they exhibit a super poissonian statistics in one of the modes and subpoissonian for the other. The transition connecting with the most populated atomic state has a super-Poissonian distribution, while the transition that connects the two less populated atomic states has a sub-Poissonian distribution. So, it seems that fields with classical statistics favor atomic transitions.

For the case in which the initial atomic state is $\ket{3_A}$ and $\tfrac{1}{\sqrt{3}}\left(\ket{1_A}+\ket{2_A}+\ket{3_A}\right)$, the statistics of the photons with frequencies $\Omega_1$ and $\Omega_2$ have the same behavior, because one has $\mathcal{P}_1(t)=\mathcal{P}_2(t)$ (Figure \ref{fig:third}). Although the linear combination for the initial atomic state occupy a large area in phase space, it is important to remark that the photons have the behavior of classical fileds since the Mandel parameter is always super-Poissonian for any value of $t$.  

At the right of Figure~\ref{fig:QM}, the behavior of the Mandel Coefficient is displayed, when the Raman conditions are considered. For the case \textit{Raman-I} the dynamics of $\mathrm{QM}_2$ remains near its initial Poisson statistics distribution function due to its weak coupling with the atom. In contrast, the dynamics for the case \textit{Raman-II} is stabilized with two different statistics super and sub-poissonian ditribution functions. 

\begin{figure}[ht]
\begin{subfigure}{0.47\textwidth}
    \includegraphics[width=\textwidth]{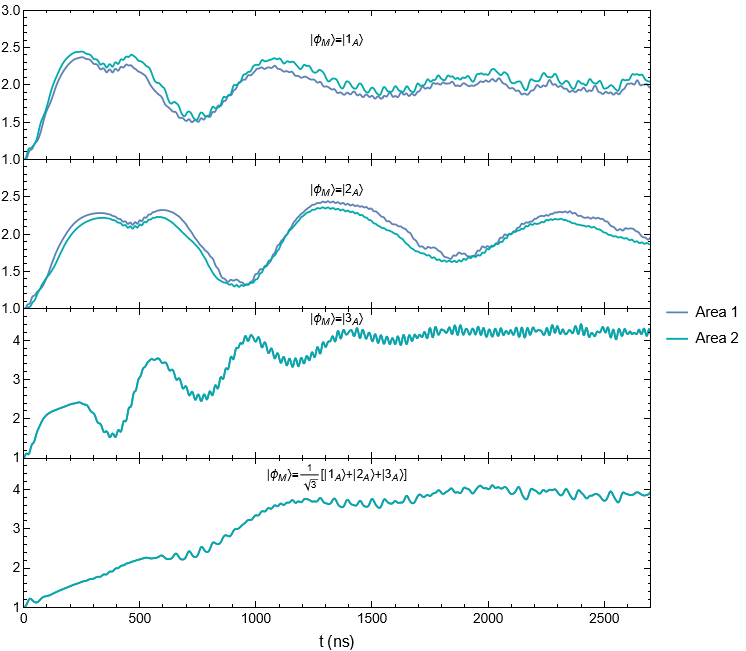}
\caption{}\label{fig:AreaEspacioFase_a}
\end{subfigure}
\hfill
\begin{subfigure}{0.49\textwidth}
    \includegraphics[width=\textwidth]{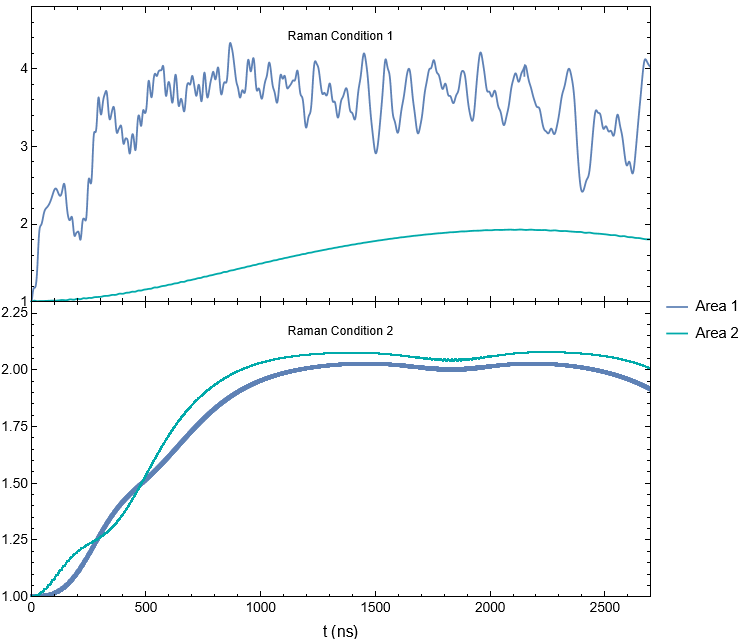}
\caption{}
 \label{fig:AreaEspacioFase_b}
\end{subfigure}
\caption{Phase-Space Area occupeid by the photons with frequency $\Omega_1$ (Area 1) and frequency $\Omega_2$ (Area 2), as a function of time. In a) the behavior of function ${\cal A} = 1/{\cal M}^{(2)}$, where ${\cal M}^{(2)}$ is defined in\eqref{second_momentum}, for the different initial states \eqref{initial_states} with its fixed conditions for the modes field. In b) the behavior of the function when the system evolves with Raman Conditions. }
\label{fig:AreaEspacioFase}
\end{figure}

The field´s behavior can be studied with the phase-space area,  ${\cal A} = 1/{\cal M}^{(2)}$, where ${\cal M}^{(2)}$ is defined in\eqref{second_momentum}. Figure \ref{fig:AreaEspacioFase} shows how  ${\cal A}$ changes because of the interaction with the atom.  Due to the initial coherent state in both fields, the initial area takes the value of 1 in all the cases presented. After the atomic interaction, it increases its value. The bigger the value gets, the more quantum the state is. 
Figure \ref{fig:AreaEspacioFase_a} is the evolution of $\cal{A}$ for the different initial atomic states. For the case when $\ket{\phi_M}=\ket{1_A}$ and $\ket{\phi_M}=\ket{2_A}$ the area oscillates around 2.5 and 1.5, but the behavior shows that it won't increase more with time. This is not the case for initial states with $\ket{\phi_M}=\ket{3_A}$ and $\ket{\phi_M}=\tfrac{1}{\sqrt{3}}[\ket{1_A}+\ket{2_A}+\ket{3_A}]$. For both cases, the phase space area increases up to a value of 4, where it stabilizes. Figure \ref{fig:AreaEspacioFase_b} shows $\cal{A}$ when Raman conditions are considered. For {\it Raman-I}, $\cal{A}$ increases near 4, but with {\it Raman-II}, $\cal{A}$ gets stabilized around 2, which is the phase space area of a cat state.   

\subsection{Dynamic entanglement properties}

For different initial states of the two-mode electromagnetic field we are going to measure the entanglement properties between the atom and the field parts of the system or between the modes of the field in the cavity. Besides of the calculation of the mutual information that they have in common. 

An state acting on a finite-dimensional Hilbert space ${\cal H}_{AB}={\cal H}_A \otimes {\cal H}_B$ is separable if the state $\rho^{AB}$ can be written in the form
\begin{equation*}
\rho^{AB} = \sum_k \, p_k \, \rho^{A}_k \otimes \rho^{B}_k \, ,
\end{equation*}
with $p_k\geq 0$, otherwise the state is entangled \cite{Guehne2009}.  A pure state $|\psi\rangle \in {\cal H}_{AB}$ is separable if it can be written as a product state, $| \psi\rangle = |\phi^A\rangle \otimes | \phi^B \rangle$, that is, if either of its reduced density matrices describes a pure state. Then, It is well known that the entanglement properties for bipartite systems can be measured by calculating the linear entropy $S_L$ or the von Neumann entropy $S_{VN}$ of the reduced density matrix $\rho^A$ or equivalently $\rho^B$,
\begin{equation}
S_L = 1- {\rm Tr}(\rho^2) \, , \quad S_{VN} = - {\rm Tr} (\rho \, \ln \, \rho) \, , \label{entropy}
\end{equation}
where $\rho=\rho^A$ or $\rho=\rho^B$. Notice that an entangled state gives more information about the total system than over the subsystems, implying the following entropic inequalities~\cite{Araki1970, Horodecki2001},
\begin{equation}
S(\rho^A) \leq S^{AB} , \qquad S(\rho^B) \leq S^{AB} \, , \quad |S^{A} -S^{B}| \leq S^{AB} \leq | S^{A} +S^{B}| \, ,
\end{equation}
which must be satisfied for the linear and von Neumann entropies.

The positive partial transposition (PPT) criterion establishes that a separable state remains a positive operator if it is subjected to partial transposition operation. The density matrix of the system in a product basis state can be written in the form
\begin{equation}
\rho _{j \, m, \, k\,  n} = \langle j  \, m \, | \rho  | k\,  n \, \rangle \, ,
\end{equation}
then the partial transposition operation with respect to the subsystem $B$ is defined by
\begin{equation}
\rho^{T_B} _{j \, m, k\,  n} = \rho _{j \, n, \, k\,  m}  \, .
\end{equation}
If one has a separable state then $\rho^{T_B} _{j \, m, k\,  n}\geq 0$ while $\rho^{T_B} _{j \, m, k\,  n}\geq 0$ does not guarantee that the state is separable for larger dimensions $n$ of the density matrices, $n>3$ \cite{Guehne2009}. 

The mutual quantum information function $I(\rho^A, \rho^B)$ is defined by
\begin{equation}
I(\rho^A, \rho^B) = S^A + S^B - S^{AB} \, ,
\end{equation}
whose dynamic behavior can be calculated for the two-modes of the electormagnetic field or for the mutual information of the atom and field sectors.

\begin{figure}[h!]
\begin{subfigure}{0.47\textwidth}
    \includegraphics[width=\textwidth]{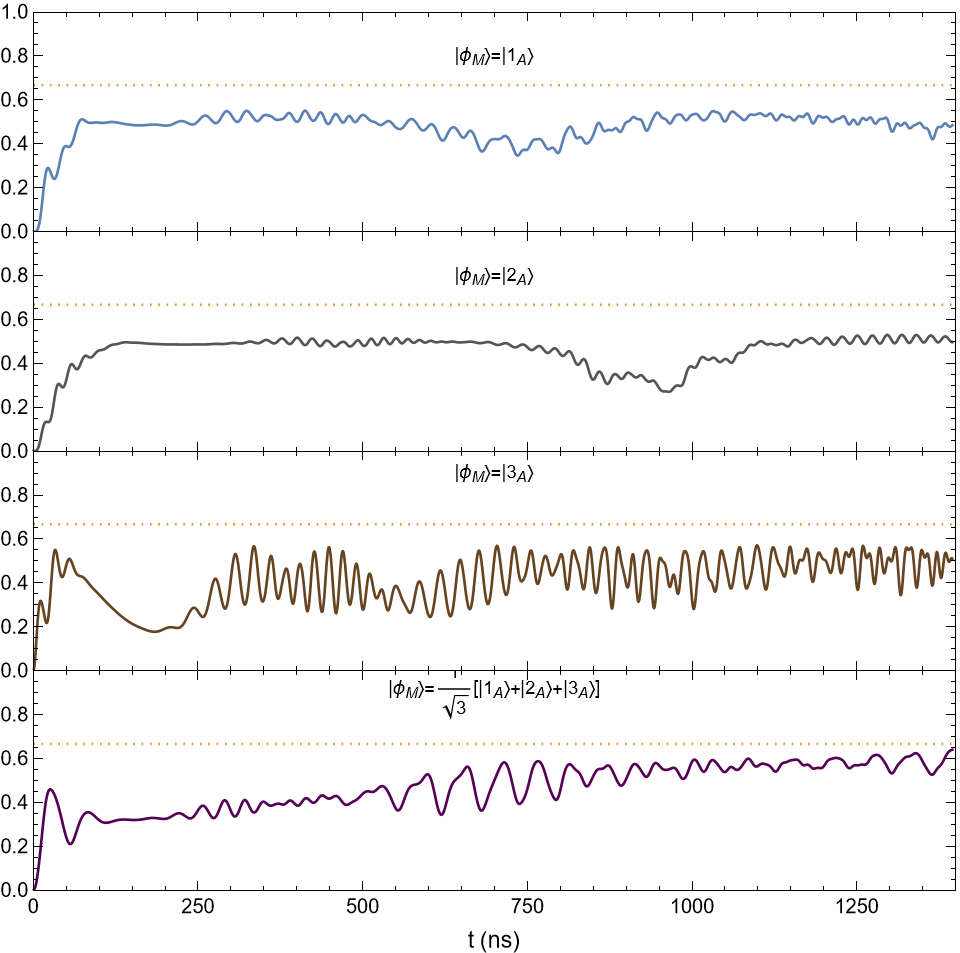}
    \caption{}
    \label{fig:SL_a}
\end{subfigure}
\hfill
\begin{subfigure}{0.49\textwidth}
    \includegraphics[width=\textwidth]{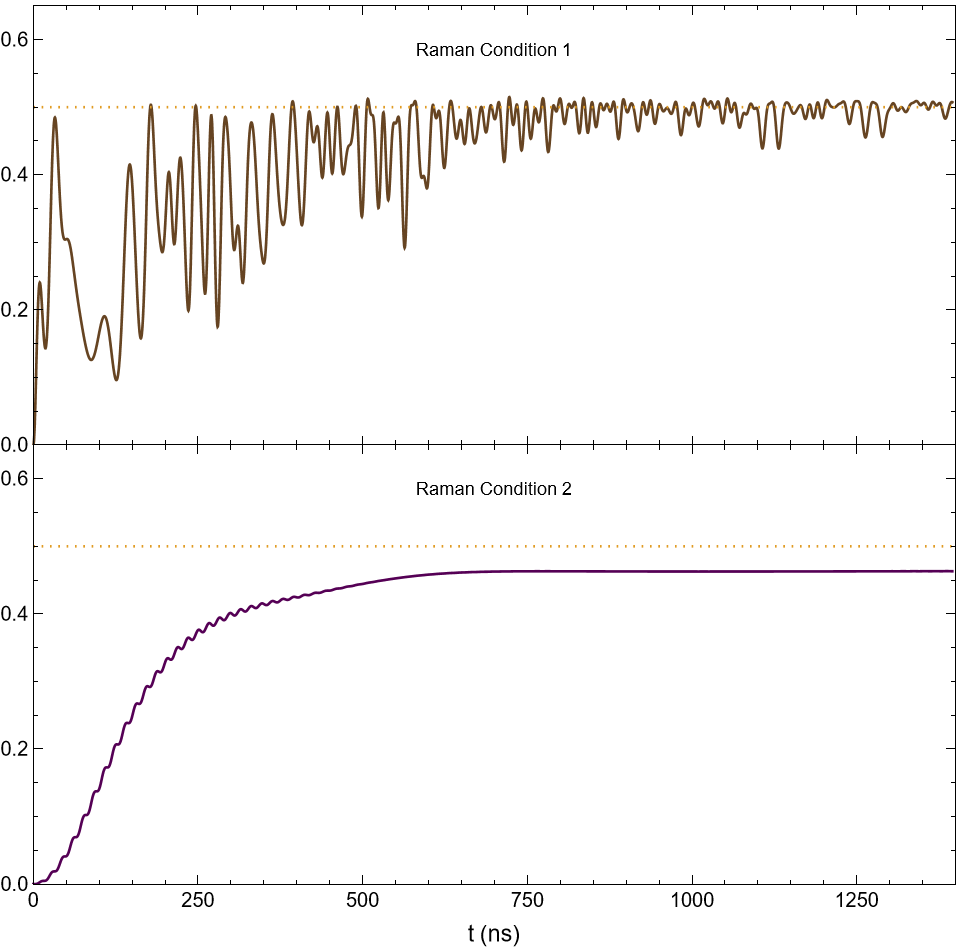}
    \caption{}
    \label{fig:SL_b}
\end{subfigure}
\caption{Linear Entropy \eqref{entropy}  between matter and field subsystems, as a function of time. In a) a measurement of the entanglement matter-field of the system when it evolves under the initial states \eqref{initial_states}, and in b) when it is under the initial Raman Conditions.  }
\label{fig:SL}
\end{figure}

Figure \ref{fig:SL} shows a measurement of the entanglement matter field of the system through the linear entropy function. For a system of three levels, the maximun entropy that the system can get is of 2/3.  In Figure \ref{fig:SL_a}, the system does not go to a maximum entanglement with no initial state \eqref{initial_states}.  Figure \ref{fig:SL_b} shows the behavior of function SL when the system evolves under the Raman Conditions. As expected, since one atomic transition is suppressed, the maximum entropy that the system can have is 1/2.

\section{Summary and Conclusions}
Experimentally, alkaline atoms can be treated as effective systems of three levels when their hyperfine levels are considered. In this work, we applied experimental data to a generalized Jaynes-Cummings Hamiltonian of two field modes \eqref{eq:Paper-001}, to model the evolution of the entanglement between the matter and both fields.

We have determined the dressed states of the system and established the properties of the dark states. By means of the constants of the motion $\hat{M}_1$ and $\hat{M}_2$ from the nine possible states for the atom-field system \eqref{eq:basis_state}, one finds that only three of them are independent. 
Thus, the time operator evolution can be constructed with any of them. We use $\left\{ \ket{\chi_{1}}, \ket{\chi_2},\ket{\chi_3}\right\}$. Since all our expressions are analytical, one can determine the evolution of all the  matter and field observables in the Fourier space. 

When condition $\Delta_{13}=\Delta_{23}$ is taken, Hamiltonian \eqref{eq:Paper-001} can be  diagonalizable with simple closed expressions, that is, three dressed states \eqref{eq:Paper-016} and \eqref{eq:Paper-017}. Although the interaction basis is composed of one dark state and two light states, it is important to remember that there are always infinite dark states \eqref{eq:darkstates} that accompany the system's evolution and that the system can access them with a probability different from zero \eqref{eqdark}. 

The interaction of the atom with the fields was studied, taking into account experimental data of isotopes $^ {6}$Li and $^ {87}$Rb.  $^ {6}$Li is the alkaline atom with the smallest energy separation between its three hyperfine levels, and $^ {87}$Rb is the atom with the biggest separation. Despite this difference, we observed that the functions evaluated in this work behave very similarly for both species. Thus, we can conclude that all alkaline atoms behave with the same patterns presented in this work. The only difference will be the unit of time, defined by $1/\omega_3$ for each atom.  

For the different initial conditions considered in this work, we can conclude that the laser lights' intensities and the number of mean photons of the field affect the  system's evolution in the structure of the patterns observed. While the intensities and the mean value of photons are bigger, the oscillations structures occur faster. An important difference is observed when we change the initial atomic state. Of particular importance are the cases when $\ket{\phi_M}=\ket{3_A}$ and $\ket{\phi_M}=\frac{1}{\sqrt{3}}\left(\ket{1_A}+\ket{2_A}+\ket{3_A}\right)$ because the atomic probabilities $\mathcal{P}_1$ and $\mathcal{P}_2$ evolve with no difference, i.e,  $\mathcal{P}_1(t)=\mathcal{P}_2(t)$, (see Figure \ref{fig:third}). 
Aditionally, functions that studied the behavior of the fields, such as the Mandel coefficient and the area of the phase-space, reflect that  we can not differentiate between photons with frequency $\Omega_1$ and frequency $\Omega_2$ (see Figure \ref{fig:QM_a} and \ref{fig:AreaEspacioFase_a}). Thus, due to the same behavior of fields connecting transitions $\ket{1_A}\leftrightarrow \ket{2_A}$ and $\ket{1_A}\leftrightarrow \ket{3_A}$,  atomic states $\ket{2_A}$ and $\ket{1_A}$ have the same occupation probabilities for all time. 

When considering the Raman conditions, we observe that an atomic transition can be suppressed via two distinct mechanisms. In Raman I, a disbalance in the intensities of the laser beams, and thus in Rabi frequencies $\mu_{12}$ and $\mu_{23}$,  is taken into account; for this case, one can control which atomic level can be suppressed between $\ket{1_A}$ and $\ket{2_A}$. In Raman II, both fields have the same conditions and the same coupling to the atom, but we introduce a very large detuning, $\Delta_{13}^2>\mu_{13}^2$ and $\Delta_{13}^2>\mu_{23}^2 $; here, the occupation probability for the state $\ket{3_A}$ is nearly zero. 

Finally, the entanglement properties of the system are measured by the function linear entropy. As expected, at $t=0$, for all the cases, the system is not entangled because the initial state is separable. Once, the interaction starts, there are quantum correlation between the matter and the fields. In Figure \ref{fig:SL_a}, for large times, the linear entropy goes nearly to the maximum value for a system of three levels, 2/3. In Figure \ref{fig:SL_b}, for large times, the linear entropy takes approximately the maximum value of a two level system, 1/2. 

\acknowledgments
FJPC and AdRL acknowledge the support of the CONAHCYT project A1-S-39242. AdRL acknowledges the scholarship of the CONAHCYT. O.C is on sabbatical leave at the Granada University and thanks support from the program PASPA of DGAPA-UNAM. 

\appendix
\section{Time Evolution Operator}

The matrix elements of the evolution operator of the system in the particle Hilbert space are given in this section in terms of the number of excitations $m_1$ and $m_2$. For expressing in terms of the number of photons, the equivalence $m_1=n_1+n_2$ and $m_2=n_2+1$ need to be introduced.  

To recall how to apply operator $\hat{U}(t)$, we use the time evolution of state $\ket{\chi_1}$ as an example. 
\begin{eqnarray*}
\ket{\chi_1(t)}&=&\left\{u_{11}(m_1,m_2,t)\ket{\chi_1}\bra{\chi_1} + u_{12}(m_1,m_2,t)\ket{\chi_2}\bra{\chi_1} +  u_{13}(m_1,m_2,t)\ket{\chi_3}\bra{\chi_1}\right\}\otimes \ket{\chi_1}\\
&=& u_{11}(m_1,m_2,t)\ket{\chi_1}+  u_{12}(m_1,m_2,t) \ket{\chi_2} + u_{13}(m_1,m_2,t) \ket{\chi_3}.
\end{eqnarray*}

\begin{eqnarray*}
u_{11} & = & \mathrm{e}^{-i\frac{E_{0}}{\hbar}t}\left\{ \frac{\mu_{23}^{2}m_{2}}{\varepsilon_{2}^{2}-\frac{\Delta_{13}^{2}}{4}}+\mu_{13}^{2}\,\left(m_{1}-m_{2}+1\right)\mathrm{e}^{-\imath\frac{\Delta_{13}}{2}t}\left[\frac{\mathrm{e}^{\imath\varepsilon_{2}t}}{2\,\varepsilon_{2}^{2}-\Delta_{13}\,\varepsilon_{2}}+\frac{\mathrm{e}^{-\imath\varepsilon_{2}t}}{2\,\varepsilon_{2}^{2}+\Delta_{13}\,\varepsilon_{2}}\right]\right\} \, , \\
u_{12} & = & \mathrm{e}^{-\imath\frac{E_{0}}{\hbar}t}\left\{ -\frac{\mu_{13}\mu_{23}\sqrt{m_{2}\left(m_{1}-m_{2}+1\right)}}{\varepsilon_{2}^{2}-\frac{\Delta_{13}^{2}}{4}}+\mu_{13}\mu_{23}\sqrt{m_{2}\left(m_{1}-m_{2}+1\right)}\mathrm{e}^{-\imath\frac{\Delta_{13}}{2}t}\left[\frac{\mathrm{e}^{\imath\varepsilon_{2}t}}{2\,\varepsilon_{2}^{2}-\Delta_{13}\,\varepsilon_{2}}+\frac{\mathrm{e}^{-\imath\varepsilon_{2}t}}{2\,\varepsilon_{2}^{2}+\Delta_{13}\,\varepsilon_{2}}\right]\right\} \, ,\\
u_{13} & = & \mathrm{e}^{-\imath\left(\frac{E_{0}}{\hbar}+\frac{\Delta_{13}}{2}\right)t}\mu_{13}\,\sqrt{m_{1}-m_{2}+1}\left\{ \frac{\mathrm{e}^{\imath\varepsilon_{2}t}}{2\,\varepsilon_{2}^{2}-\Delta_{13}\,\varepsilon_{2}}\left[\frac{\Delta_{13}}{2}-\varepsilon_{2}\right]+\frac{\mathrm{e}^{-\imath\varepsilon_{2}t}}{2\,\varepsilon_{2}^{2}+\Delta_{13}\,\varepsilon_{2}}\left[\frac{\Delta_{13}}{2}+\varepsilon_{2}\right]\right\} \, ,\\
u_{21} & = & \mathrm{e}^{-\imath\frac{E_{0}}{\hbar}t}\left\{ -\frac{\mu_{13}\mu_{23}\sqrt{m_{2}\left(m_{1}-m_{2}+1\right)}}{\varepsilon_{2}^{2}-\frac{\Delta_{13}^{2}}{4}}+\mathrm{e}^{-\imath\frac{\Delta_{13}}{2}t}\mu_{13}\mu_{23}\sqrt{m_{2}\left(m_{1}-m_{2}+1\right)}\left[\frac{\mathrm{e}^{\imath\varepsilon_{2}t}}{2\,\varepsilon_{2}^{2}-\Delta_{13}\,\varepsilon_{2}}+\frac{\mathrm{e}^{-\imath\varepsilon_{2}t}}{2\,\varepsilon_{2}^{2}+\Delta_{13}\,\varepsilon_{2}}\right]\right\} \, ,\\
u_{22} & = & \mathrm{e}^{-\imath\frac{E_{0}}{\hbar}t}\left\{ \frac{\mu_{13}^{2}\left(m_{1}-m_{2}+1\right)}{\varepsilon_{2}^{2}-\frac{\Delta_{13}^{2}}{4}}+\mathrm{e}^{-\imath\frac{\Delta_{13}}{2}t}\mu_{23}^{2}m_{2}\left[\frac{\mathrm{e}^{\imath\varepsilon_{2}t}}{2\,\varepsilon_{2}^{2}-\Delta_{13}\,\varepsilon_{2}}+\frac{\mathrm{e}^{-\imath\varepsilon_{2}t}}{2\,\varepsilon_{2}^{2}+\Delta_{13}\,\varepsilon_{2}}\right]\right\} \, , \\
u_{23} & = & -\imath\mathrm{e}^{-\imath\left(\frac{E_{0}}{\hbar}+\frac{\Delta_{13}}{2}\right)t}\,\frac{\mu_{23}\sqrt{m_{2}}}{\varepsilon_{2}}\sin\varepsilon_{2}t \, ,\\
u_{31} & = & -\imath\mathrm{e}^{-\imath\left(\frac{E_{0}}{\hbar}+\frac{\Delta_{13}}{2}\right)t}\,\frac{\mu_{13}\sqrt{m_{1}-m_{2}+1}}{\varepsilon_{2}}\sin\varepsilon_{2}t \, , \\
u_{32} & = & -\imath\mathrm{e}^{-\imath\left(\frac{E_{0}}{\hbar}+\frac{\Delta_{13}}{2}\right)t}\,\frac{\mu_{23}\sqrt{m_{2}}}{\varepsilon_{2}}\sin\varepsilon_{2}t \, , \\
u_{33} & = & \mathrm{e}^{-\imath\left(\frac{E_{0}}{\hbar}+\frac{\Delta_{13}}{2}\right)t}\left\{ \cos\varepsilon_{2}t-\imath\frac{\Delta_{13}}{2\varepsilon_{2}}\sin\varepsilon_{2}t\right\} \, .
\end{eqnarray*}

\bibliographystyle{apsrev4-1}
\bibliography{bib3levels}

\begin{thebibliography}{54}%
\makeatletter
\providecommand \@ifxundefined [1]{%
 \@ifx{#1\undefined}
}%
\providecommand \@ifnum [1]{%
 \ifnum #1\expandafter \@firstoftwo
 \else \expandafter \@secondoftwo
 \fi
}%
\providecommand \@ifx [1]{%
 \ifx #1\expandafter \@firstoftwo
 \else \expandafter \@secondoftwo
 \fi
}%
\providecommand \natexlab [1]{#1}%
\providecommand \enquote  [1]{``#1''}%
\providecommand \bibnamefont  [1]{#1}%
\providecommand \bibfnamefont [1]{#1}%
\providecommand \citenamefont [1]{#1}%
\providecommand \href@noop [0]{\@secondoftwo}%
\providecommand \href [0]{\begingroup \@sanitize@url \@href}%
\providecommand \@href[1]{\@@startlink{#1}\@@href}%
\providecommand \@@href[1]{\endgroup#1\@@endlink}%
\providecommand \@sanitize@url [0]{\catcode `\\12\catcode `\$12\catcode
  `\&12\catcode `\#12\catcode `\^12\catcode `\_12\catcode `\%12\relax}%
\providecommand \@@startlink[1]{}%
\providecommand \@@endlink[0]{}%
\providecommand \url  [0]{\begingroup\@sanitize@url \@url }%
\providecommand \@url [1]{\endgroup\@href {#1}{\urlprefix }}%
\providecommand \urlprefix  [0]{URL }%
\providecommand \Eprint [0]{\href }%
\providecommand \doibase [0]{http://dx.doi.org/}%
\providecommand \selectlanguage [0]{\@gobble}%
\providecommand \bibinfo  [0]{\@secondoftwo}%
\providecommand \bibfield  [0]{\@secondoftwo}%
\providecommand \translation [1]{[#1]}%
\providecommand \BibitemOpen [0]{}%
\providecommand \bibitemStop [0]{}%
\providecommand \bibitemNoStop [0]{.\EOS\space}%
\providecommand \EOS [0]{\spacefactor3000\relax}%
\providecommand \BibitemShut  [1]{\csname bibitem#1\endcsname}%
\let\auto@bib@innerbib\@empty
\bibitem [{\citenamefont {Einstein}\ \emph {et~al.}(1935)\citenamefont
  {Einstein}, \citenamefont {Podolsky},\ and\ \citenamefont
  {Rosen}}]{Einstein1935}%
  \BibitemOpen
  \bibfield  {author} {\bibinfo {author} {\bibfnamefont {A.}~\bibnamefont
  {Einstein}}, \bibinfo {author} {\bibfnamefont {B.}~\bibnamefont {Podolsky}},
  \ and\ \bibinfo {author} {\bibfnamefont {N.}~\bibnamefont {Rosen}},\ }\href
  {\doibase 10.1103/PhysRev.47.777} {\bibfield  {journal} {\bibinfo  {journal}
  {Phys. Rev.}\ }\textbf {\bibinfo {volume} {47}},\ \bibinfo {pages} {777}
  (\bibinfo {year} {1935})}\BibitemShut {NoStop}%
\bibitem [{\citenamefont {Schrödinger}(1935)}]{Schroedinger1935}%
  \BibitemOpen
  \bibfield  {author} {\bibinfo {author} {\bibfnamefont {E.}~\bibnamefont
  {Schrödinger}},\ }\href {\doibase 10.1017/S0305004100013554} {\bibfield
  {journal} {\bibinfo  {journal} {Math. Proc. Cambridge Philos. Soc.}\ }\textbf
  {\bibinfo {volume} {31}},\ \bibinfo {pages} {555–563} (\bibinfo {year}
  {1935})}\BibitemShut {NoStop}%
\bibitem [{\citenamefont {Bell}\ and\ \citenamefont {Aspect}(2004)}]{Bell2004}%
  \BibitemOpen
  \bibfield  {author} {\bibinfo {author} {\bibfnamefont {J.~S.}\ \bibnamefont
  {Bell}}\ and\ \bibinfo {author} {\bibfnamefont {A.}~\bibnamefont {Aspect}},\
  }\href@noop {} {\emph {\bibinfo {title} {Speakable and Unspeakable in Quantum
  Mechanics: Collected Papers on Quantum Philosophy}}},\ \bibinfo {edition}
  {2nd}\ ed.\ (\bibinfo  {publisher} {Cambridge University Press},\ \bibinfo
  {year} {2004})\BibitemShut {NoStop}%
\bibitem [{\citenamefont {Clauser}\ \emph {et~al.}(1969)\citenamefont
  {Clauser}, \citenamefont {Horne}, \citenamefont {Shimony},\ and\
  \citenamefont {Holt}}]{Clauser1969}%
  \BibitemOpen
  \bibfield  {author} {\bibinfo {author} {\bibfnamefont {J.~F.}\ \bibnamefont
  {Clauser}}, \bibinfo {author} {\bibfnamefont {M.~A.}\ \bibnamefont {Horne}},
  \bibinfo {author} {\bibfnamefont {A.}~\bibnamefont {Shimony}}, \ and\
  \bibinfo {author} {\bibfnamefont {R.~A.}\ \bibnamefont {Holt}},\ }\href
  {\doibase 10.1103/PhysRevLett.23.880} {\bibfield  {journal} {\bibinfo
  {journal} {Phys. Rev. Lett.}\ }\textbf {\bibinfo {volume} {23}},\ \bibinfo
  {pages} {880} (\bibinfo {year} {1969})}\BibitemShut {NoStop}%
\bibitem [{\citenamefont {Aspect}\ \emph {et~al.}(1982)\citenamefont {Aspect},
  \citenamefont {Grangier},\ and\ \citenamefont {Roger}}]{Aspect1982}%
  \BibitemOpen
  \bibfield  {author} {\bibinfo {author} {\bibfnamefont {A.}~\bibnamefont
  {Aspect}}, \bibinfo {author} {\bibfnamefont {P.}~\bibnamefont {Grangier}}, \
  and\ \bibinfo {author} {\bibfnamefont {G.}~\bibnamefont {Roger}},\ }\href
  {\doibase 10.1103/PhysRevLett.49.91} {\bibfield  {journal} {\bibinfo
  {journal} {Phys. Rev. Lett.}\ }\textbf {\bibinfo {volume} {49}},\ \bibinfo
  {pages} {91} (\bibinfo {year} {1982})}\BibitemShut {NoStop}%
\bibitem [{\citenamefont {Gühne}\ and\ \citenamefont
  {Tóth}(2009)}]{Guehne2009}%
  \BibitemOpen
  \bibfield  {author} {\bibinfo {author} {\bibfnamefont {O.}~\bibnamefont
  {Gühne}}\ and\ \bibinfo {author} {\bibfnamefont {G.}~\bibnamefont {Tóth}},\
  }\href {\doibase https://doi.org/10.1016/j.physrep.2009.02.004} {\bibfield
  {journal} {\bibinfo  {journal} {Phys. Rep.}\ }\textbf {\bibinfo {volume}
  {474}},\ \bibinfo {pages} {1} (\bibinfo {year} {2009})},\ \Eprint
  {http://arxiv.org/abs/and references there in} {and references there in}
  \BibitemShut {NoStop}%
\bibitem [{\citenamefont {Barnett}\ \emph {et~al.}(2017)\citenamefont
  {Barnett}, \citenamefont {Beige}, \citenamefont {Ekert}, \citenamefont
  {Garraway}, \citenamefont {Keitel}, \citenamefont {Kendon}, \citenamefont
  {Lein}, \citenamefont {Milburn}, \citenamefont {Moya-Cessa}, \citenamefont
  {Murao}, \citenamefont {Pachos}, \citenamefont {Palma}, \citenamefont
  {Paspalakis}, \citenamefont {Phoenix}, \citenamefont {Piraux}, \citenamefont
  {Plenio}, \citenamefont {Sanders}, \citenamefont {Twamley}, \citenamefont
  {Vidiella-Barranco},\ and\ \citenamefont {Kim}}]{Barnett2017}%
  \BibitemOpen
  \bibfield  {author} {\bibinfo {author} {\bibfnamefont {S.~M.}\ \bibnamefont
  {Barnett}}, \bibinfo {author} {\bibfnamefont {A.}~\bibnamefont {Beige}},
  \bibinfo {author} {\bibfnamefont {A.}~\bibnamefont {Ekert}}, \bibinfo
  {author} {\bibfnamefont {B.~M.}\ \bibnamefont {Garraway}}, \bibinfo {author}
  {\bibfnamefont {C.~H.}\ \bibnamefont {Keitel}}, \bibinfo {author}
  {\bibfnamefont {V.}~\bibnamefont {Kendon}}, \bibinfo {author} {\bibfnamefont
  {M.}~\bibnamefont {Lein}}, \bibinfo {author} {\bibfnamefont {G.~J.}\
  \bibnamefont {Milburn}}, \bibinfo {author} {\bibfnamefont {H.~M.}\
  \bibnamefont {Moya-Cessa}}, \bibinfo {author} {\bibfnamefont
  {M.}~\bibnamefont {Murao}}, \bibinfo {author} {\bibfnamefont {J.~K.}\
  \bibnamefont {Pachos}}, \bibinfo {author} {\bibfnamefont {G.~M.}\
  \bibnamefont {Palma}}, \bibinfo {author} {\bibfnamefont {E.}~\bibnamefont
  {Paspalakis}}, \bibinfo {author} {\bibfnamefont {S.~J.}\ \bibnamefont
  {Phoenix}}, \bibinfo {author} {\bibfnamefont {B.}~\bibnamefont {Piraux}},
  \bibinfo {author} {\bibfnamefont {M.~B.}\ \bibnamefont {Plenio}}, \bibinfo
  {author} {\bibfnamefont {B.~C.}\ \bibnamefont {Sanders}}, \bibinfo {author}
  {\bibfnamefont {J.}~\bibnamefont {Twamley}}, \bibinfo {author} {\bibfnamefont
  {A.}~\bibnamefont {Vidiella-Barranco}}, \ and\ \bibinfo {author}
  {\bibfnamefont {M.}~\bibnamefont {Kim}},\ }\href {\doibase
  https://doi.org/10.1016/j.pquantelec.2017.07.002} {\bibfield  {journal}
  {\bibinfo  {journal} {Progress in Quantum Electronics}\ }\textbf {\bibinfo
  {volume} {54}},\ \bibinfo {pages} {19} (\bibinfo {year} {2017})},\ \bibinfo
  {note} {special issue in honor of the 70th birthday of Professor Sir Peter
  Knight FRS}\BibitemShut {NoStop}%
\bibitem [{\citenamefont {Barnett}\ and\ \citenamefont
  {Phoenix}(1989)}]{Barnett1989}%
  \BibitemOpen
  \bibfield  {author} {\bibinfo {author} {\bibfnamefont {S.~M.}\ \bibnamefont
  {Barnett}}\ and\ \bibinfo {author} {\bibfnamefont {S.~J.~D.}\ \bibnamefont
  {Phoenix}},\ }\href {\doibase 10.1103/PhysRevA.40.2404} {\bibfield  {journal}
  {\bibinfo  {journal} {Phys. Rev. A}\ }\textbf {\bibinfo {volume} {40}},\
  \bibinfo {pages} {2404} (\bibinfo {year} {1989})}\BibitemShut {NoStop}%
\bibitem [{\citenamefont {Barnett}(2009)}]{Barnett2009}%
  \BibitemOpen
  \bibfield  {author} {\bibinfo {author} {\bibfnamefont {S.}~\bibnamefont
  {Barnett}},\ }\href@noop {} {\emph {\bibinfo {title} {Quantum
  Information}}},\ Oxford Master Series in Physics\ (\bibinfo  {publisher} {OUP
  Oxford},\ \bibinfo {year} {2009})\BibitemShut {NoStop}%
\bibitem [{\citenamefont {Haroche}\ and\ \citenamefont
  {Raimond}(2013)}]{Haroche2013}%
  \BibitemOpen
  \bibfield  {author} {\bibinfo {author} {\bibfnamefont {S.}~\bibnamefont
  {Haroche}}\ and\ \bibinfo {author} {\bibfnamefont {J.}~\bibnamefont
  {Raimond}},\ }\href@noop {} {\emph {\bibinfo {title} {Exploring the Quantum:
  Atoms, Cavities, and Photons}}},\ Oxford Graduate Texts\ (\bibinfo
  {publisher} {OUP Oxford},\ \bibinfo {year} {2013})\BibitemShut {NoStop}%
\bibitem [{\citenamefont {Pike}\ and\ \citenamefont {Sarkar}(1986)}]{Pike1986}%
  \BibitemOpen
  \bibfield  {author} {\bibinfo {author} {\bibfnamefont {E.}~\bibnamefont
  {Pike}}\ and\ \bibinfo {author} {\bibfnamefont {S.}~\bibnamefont {Sarkar}},\
  }\href@noop {} {\emph {\bibinfo {title} {Frontiers in Quantum Optics,}}}\
  (\bibinfo  {publisher} {Taylor \& Francis},\ \bibinfo {year}
  {1986})\BibitemShut {NoStop}%
\bibitem [{\citenamefont {Yoo}\ and\ \citenamefont {Eberly}(1985)}]{Yoo1985}%
  \BibitemOpen
  \bibfield  {author} {\bibinfo {author} {\bibfnamefont {H.-I.}\ \bibnamefont
  {Yoo}}\ and\ \bibinfo {author} {\bibfnamefont {J.}~\bibnamefont {Eberly}},\
  }\href {\doibase https://doi.org/10.1016/0370-1573(85)90015-8} {\bibfield
  {journal} {\bibinfo  {journal} {Phys. Rep.}\ }\textbf {\bibinfo {volume}
  {118}},\ \bibinfo {pages} {239} (\bibinfo {year} {1985})}\BibitemShut
  {NoStop}%
\bibitem [{\citenamefont {Kaluzny}\ \emph {et~al.}(1983)\citenamefont
  {Kaluzny}, \citenamefont {Goy}, \citenamefont {Gross}, \citenamefont
  {Raimond},\ and\ \citenamefont {Haroche}}]{Kaluzny1983}%
  \BibitemOpen
  \bibfield  {author} {\bibinfo {author} {\bibfnamefont {Y.}~\bibnamefont
  {Kaluzny}}, \bibinfo {author} {\bibfnamefont {P.}~\bibnamefont {Goy}},
  \bibinfo {author} {\bibfnamefont {M.}~\bibnamefont {Gross}}, \bibinfo
  {author} {\bibfnamefont {J.~M.}\ \bibnamefont {Raimond}}, \ and\ \bibinfo
  {author} {\bibfnamefont {S.}~\bibnamefont {Haroche}},\ }\href {\doibase
  10.1103/PhysRevLett.51.1175} {\bibfield  {journal} {\bibinfo  {journal}
  {Phys. Rev. Lett.}\ }\textbf {\bibinfo {volume} {51}},\ \bibinfo {pages}
  {1175} (\bibinfo {year} {1983})}\BibitemShut {NoStop}%
\bibitem [{\citenamefont {Rempe}\ \emph {et~al.}(1987)\citenamefont {Rempe},
  \citenamefont {Walther},\ and\ \citenamefont {Klein}}]{Rempe1987}%
  \BibitemOpen
  \bibfield  {author} {\bibinfo {author} {\bibfnamefont {G.}~\bibnamefont
  {Rempe}}, \bibinfo {author} {\bibfnamefont {H.}~\bibnamefont {Walther}}, \
  and\ \bibinfo {author} {\bibfnamefont {N.}~\bibnamefont {Klein}},\ }\href
  {\doibase 10.1103/PhysRevLett.58.353} {\bibfield  {journal} {\bibinfo
  {journal} {Phys. Rev. Lett.}\ }\textbf {\bibinfo {volume} {58}},\ \bibinfo
  {pages} {353} (\bibinfo {year} {1987})}\BibitemShut {NoStop}%
\bibitem [{\citenamefont {Raizen}\ \emph {et~al.}(1989)\citenamefont {Raizen},
  \citenamefont {Thompson}, \citenamefont {Brecha}, \citenamefont {Kimble},\
  and\ \citenamefont {Carmichael}}]{Raizen1989}%
  \BibitemOpen
  \bibfield  {author} {\bibinfo {author} {\bibfnamefont {M.~G.}\ \bibnamefont
  {Raizen}}, \bibinfo {author} {\bibfnamefont {R.~J.}\ \bibnamefont
  {Thompson}}, \bibinfo {author} {\bibfnamefont {R.~J.}\ \bibnamefont
  {Brecha}}, \bibinfo {author} {\bibfnamefont {H.~J.}\ \bibnamefont {Kimble}},
  \ and\ \bibinfo {author} {\bibfnamefont {H.~J.}\ \bibnamefont {Carmichael}},\
  }\href {\doibase 10.1103/PhysRevLett.63.240} {\bibfield  {journal} {\bibinfo
  {journal} {Phys. Rev. Lett.}\ }\textbf {\bibinfo {volume} {63}},\ \bibinfo
  {pages} {240} (\bibinfo {year} {1989})}\BibitemShut {NoStop}%
\bibitem [{\citenamefont {Jaynes}\ and\ \citenamefont
  {Cummings}(1963)}]{Jaynes1963}%
  \BibitemOpen
  \bibfield  {author} {\bibinfo {author} {\bibfnamefont {E.}~\bibnamefont
  {Jaynes}}\ and\ \bibinfo {author} {\bibfnamefont {F.}~\bibnamefont
  {Cummings}},\ }\href {\doibase 10.1109/PROC.1963.1664} {\bibfield  {journal}
  {\bibinfo  {journal} {Proc. IEEE}\ }\textbf {\bibinfo {volume} {51}},\
  \bibinfo {pages} {89} (\bibinfo {year} {1963})}\BibitemShut {NoStop}%
\bibitem [{\citenamefont {Eberly}\ \emph {et~al.}(1980)\citenamefont {Eberly},
  \citenamefont {Narozhny},\ and\ \citenamefont
  {Sanchez-Mondragon}}]{Eberly1980}%
  \BibitemOpen
  \bibfield  {author} {\bibinfo {author} {\bibfnamefont {J.~H.}\ \bibnamefont
  {Eberly}}, \bibinfo {author} {\bibfnamefont {N.~B.}\ \bibnamefont
  {Narozhny}}, \ and\ \bibinfo {author} {\bibfnamefont {J.~J.}\ \bibnamefont
  {Sanchez-Mondragon}},\ }\href {\doibase 10.1103/PhysRevLett.44.1323}
  {\bibfield  {journal} {\bibinfo  {journal} {Phys. Rev. Lett.}\ }\textbf
  {\bibinfo {volume} {44}},\ \bibinfo {pages} {1323} (\bibinfo {year}
  {1980})}\BibitemShut {NoStop}%
\bibitem [{\citenamefont {Narozhny}\ \emph {et~al.}(1981)\citenamefont
  {Narozhny}, \citenamefont {Sanchez-Mondragon},\ and\ \citenamefont
  {Eberly}}]{Narozhny1981}%
  \BibitemOpen
  \bibfield  {author} {\bibinfo {author} {\bibfnamefont {N.~B.}\ \bibnamefont
  {Narozhny}}, \bibinfo {author} {\bibfnamefont {J.~J.}\ \bibnamefont
  {Sanchez-Mondragon}}, \ and\ \bibinfo {author} {\bibfnamefont {J.~H.}\
  \bibnamefont {Eberly}},\ }\href {\doibase 10.1103/PhysRevA.23.236} {\bibfield
   {journal} {\bibinfo  {journal} {Phys. Rev. A}\ }\textbf {\bibinfo {volume}
  {23}},\ \bibinfo {pages} {236} (\bibinfo {year} {1981})}\BibitemShut
  {NoStop}%
\bibitem [{\citenamefont {Meschede}\ \emph {et~al.}(1985)\citenamefont
  {Meschede}, \citenamefont {Walther},\ and\ \citenamefont
  {M\"uller}}]{Meschede1985}%
  \BibitemOpen
  \bibfield  {author} {\bibinfo {author} {\bibfnamefont {D.}~\bibnamefont
  {Meschede}}, \bibinfo {author} {\bibfnamefont {H.}~\bibnamefont {Walther}}, \
  and\ \bibinfo {author} {\bibfnamefont {G.}~\bibnamefont {M\"uller}},\ }\href
  {\doibase 10.1103/PhysRevLett.54.551} {\bibfield  {journal} {\bibinfo
  {journal} {Phys. Rev. Lett.}\ }\textbf {\bibinfo {volume} {54}},\ \bibinfo
  {pages} {551} (\bibinfo {year} {1985})}\BibitemShut {NoStop}%
\bibitem [{\citenamefont {Gerry}\ and\ \citenamefont
  {Moyer}(1988)}]{Gerry1988}%
  \BibitemOpen
  \bibfield  {author} {\bibinfo {author} {\bibfnamefont {C.~C.}\ \bibnamefont
  {Gerry}}\ and\ \bibinfo {author} {\bibfnamefont {P.~J.}\ \bibnamefont
  {Moyer}},\ }\href {\doibase 10.1103/PhysRevA.38.5665} {\bibfield  {journal}
  {\bibinfo  {journal} {Phys. Rev. A}\ }\textbf {\bibinfo {volume} {38}},\
  \bibinfo {pages} {5665} (\bibinfo {year} {1988})}\BibitemShut {NoStop}%
\bibitem [{\citenamefont {He}(1989)}]{He1989}%
  \BibitemOpen
  \bibfield  {author} {\bibinfo {author} {\bibfnamefont {L.-s.}\ \bibnamefont
  {He}},\ }\href {\doibase 10.1364/JOSAB.6.001915} {\bibfield  {journal}
  {\bibinfo  {journal} {J. Opt. Soc. Am. B}\ }\textbf {\bibinfo {volume} {6}},\
  \bibinfo {pages} {1915} (\bibinfo {year} {1989})}\BibitemShut {NoStop}%
\bibitem [{\citenamefont {Stenholm}(1973)}]{Stenholm1973}%
  \BibitemOpen
  \bibfield  {author} {\bibinfo {author} {\bibfnamefont {S.}~\bibnamefont
  {Stenholm}},\ }\href {\doibase https://doi.org/10.1016/0370-1573(73)90011-2}
  {\bibfield  {journal} {\bibinfo  {journal} {Phys. Rep.}\ }\textbf {\bibinfo
  {volume} {6}},\ \bibinfo {pages} {1} (\bibinfo {year} {1973})}\BibitemShut
  {NoStop}%
\bibitem [{\citenamefont {Jr}\ \emph {et~al.}(1986)\citenamefont {Jr},
  \citenamefont {Kien},\ and\ \citenamefont {Shumovsky}}]{Jr1986}%
  \BibitemOpen
  \bibfield  {author} {\bibinfo {author} {\bibfnamefont {N.~N.~B.}\
  \bibnamefont {Jr}}, \bibinfo {author} {\bibfnamefont {F.~L.}\ \bibnamefont
  {Kien}}, \ and\ \bibinfo {author} {\bibfnamefont {A.~S.}\ \bibnamefont
  {Shumovsky}},\ }\href {\doibase 10.1088/0305-4470/19/2/015} {\bibfield
  {journal} {\bibinfo  {journal} {J. Phys. A: Math. Gen.}\ }\textbf {\bibinfo
  {volume} {19}},\ \bibinfo {pages} {191} (\bibinfo {year} {1986})}\BibitemShut
  {NoStop}%
\bibitem [{\citenamefont {Li}\ and\ \citenamefont {Peng}(1985)}]{Li1985}%
  \BibitemOpen
  \bibfield  {author} {\bibinfo {author} {\bibfnamefont {X.-s.}\ \bibnamefont
  {Li}}\ and\ \bibinfo {author} {\bibfnamefont {Y.-n.}\ \bibnamefont {Peng}},\
  }\href {\doibase 10.1103/PhysRevA.32.1501} {\bibfield  {journal} {\bibinfo
  {journal} {Phys. Rev. A}\ }\textbf {\bibinfo {volume} {32}},\ \bibinfo
  {pages} {1501} (\bibinfo {year} {1985})}\BibitemShut {NoStop}%
\bibitem [{\citenamefont {Zhu}(1989)}]{Zhu1989}%
  \BibitemOpen
  \bibfield  {author} {\bibinfo {author} {\bibfnamefont {S.-Y.}\ \bibnamefont
  {Zhu}},\ }\href {\doibase 10.1080/09500348914550571} {\bibfield  {journal}
  {\bibinfo  {journal} {J. Mod. Optic.}\ }\textbf {\bibinfo {volume} {36}},\
  \bibinfo {pages} {499} (\bibinfo {year} {1989})}\BibitemShut {NoStop}%
\bibitem [{\citenamefont {Phoenix}\ and\ \citenamefont
  {Knight}(1988)}]{Phoenix1988}%
  \BibitemOpen
  \bibfield  {author} {\bibinfo {author} {\bibfnamefont {S.}~\bibnamefont
  {Phoenix}}\ and\ \bibinfo {author} {\bibfnamefont {P.}~\bibnamefont
  {Knight}},\ }\href {\doibase https://doi.org/10.1016/0003-4916(88)90006-1}
  {\bibfield  {journal} {\bibinfo  {journal} {Ann. Phys-New. York.}\ }\textbf
  {\bibinfo {volume} {186}},\ \bibinfo {pages} {381} (\bibinfo {year}
  {1988})}\BibitemShut {NoStop}%
\bibitem [{\citenamefont {Boukobza}\ and\ \citenamefont
  {Tannor}(2005)}]{Boukobza2005}%
  \BibitemOpen
  \bibfield  {author} {\bibinfo {author} {\bibfnamefont {E.}~\bibnamefont
  {Boukobza}}\ and\ \bibinfo {author} {\bibfnamefont {D.~J.}\ \bibnamefont
  {Tannor}},\ }\href {\doibase 10.1103/PhysRevA.71.063821} {\bibfield
  {journal} {\bibinfo  {journal} {Phys. Rev. A}\ }\textbf {\bibinfo {volume}
  {71}},\ \bibinfo {pages} {063821} (\bibinfo {year} {2005})}\BibitemShut
  {NoStop}%
\bibitem [{\citenamefont {Gerry}\ and\ \citenamefont
  {Eberly}(1990)}]{Gerry1990}%
  \BibitemOpen
  \bibfield  {author} {\bibinfo {author} {\bibfnamefont {C.~C.}\ \bibnamefont
  {Gerry}}\ and\ \bibinfo {author} {\bibfnamefont {J.~H.}\ \bibnamefont
  {Eberly}},\ }\href {\doibase 10.1103/PhysRevA.42.6805} {\bibfield  {journal}
  {\bibinfo  {journal} {Phys. Rev. A}\ }\textbf {\bibinfo {volume} {42}},\
  \bibinfo {pages} {6805} (\bibinfo {year} {1990})}\BibitemShut {NoStop}%
\bibitem [{\citenamefont {Knight}(1986)}]{Knight1986}%
  \BibitemOpen
  \bibfield  {author} {\bibinfo {author} {\bibfnamefont {P.~L.}\ \bibnamefont
  {Knight}},\ }\href {\doibase 10.1088/0031-8949/1986/T12/007} {\bibfield
  {journal} {\bibinfo  {journal} {Phys. Scripta}\ }\textbf {\bibinfo {volume}
  {1986}},\ \bibinfo {pages} {51} (\bibinfo {year} {1986})}\BibitemShut
  {NoStop}%
\bibitem [{\citenamefont {Phoenix}\ and\ \citenamefont
  {Knight}(1990)}]{Phoenix1990}%
  \BibitemOpen
  \bibfield  {author} {\bibinfo {author} {\bibfnamefont {S.~J.~D.}\
  \bibnamefont {Phoenix}}\ and\ \bibinfo {author} {\bibfnamefont {P.~L.}\
  \bibnamefont {Knight}},\ }\href {\doibase 10.1364/JOSAB.7.000116} {\bibfield
  {journal} {\bibinfo  {journal} {J. Opt. Soc. Am. B}\ }\textbf {\bibinfo
  {volume} {7}},\ \bibinfo {pages} {116} (\bibinfo {year} {1990})}\BibitemShut
  {NoStop}%
\bibitem [{\citenamefont {Gou}(1989)}]{Gou1989}%
  \BibitemOpen
  \bibfield  {author} {\bibinfo {author} {\bibfnamefont {S.-C.}\ \bibnamefont
  {Gou}},\ }\href {\doibase 10.1103/PhysRevA.40.5116} {\bibfield  {journal}
  {\bibinfo  {journal} {Phys. Rev. A}\ }\textbf {\bibinfo {volume} {40}},\
  \bibinfo {pages} {5116} (\bibinfo {year} {1989})}\BibitemShut {NoStop}%
\bibitem [{\citenamefont {Bogolubov}\ \emph {et~al.}(1986)\citenamefont
  {Bogolubov}, \citenamefont {Le~Kien},\ and\ \citenamefont
  {Shumovsky}}]{Bogolubov1986}%
  \BibitemOpen
  \bibfield  {author} {\bibinfo {author} {\bibfnamefont {N.}~\bibnamefont
  {Bogolubov}}, \bibinfo {author} {\bibfnamefont {F.}~\bibnamefont {Le~Kien}},
  \ and\ \bibinfo {author} {\bibfnamefont {A.}~\bibnamefont {Shumovsky}},\
  }\href {\doibase 10.1051/jphys:01986004703042700} {\bibfield  {journal}
  {\bibinfo  {journal} {J. Phys. France}\ }\textbf {\bibinfo {volume} {47}},\
  \bibinfo {pages} {427} (\bibinfo {year} {1986})}\BibitemShut {NoStop}%
\bibitem [{\citenamefont {Bogolubov}\ \emph {et~al.}(1987)\citenamefont
  {Bogolubov}, \citenamefont {Kien},\ and\ \citenamefont
  {Shumovsky}}]{Bogolubov1987}%
  \BibitemOpen
  \bibfield  {author} {\bibinfo {author} {\bibfnamefont {N.~N.}\ \bibnamefont
  {Bogolubov}}, \bibinfo {author} {\bibfnamefont {F.~L.}\ \bibnamefont {Kien}},
  \ and\ \bibinfo {author} {\bibfnamefont {A.~S.}\ \bibnamefont {Shumovsky}},\
  }\href {\doibase 10.1209/0295-5075/4/3/005} {\bibfield  {journal} {\bibinfo
  {journal} {Europhys. Lett.}\ }\textbf {\bibinfo {volume} {4}},\ \bibinfo
  {pages} {281} (\bibinfo {year} {1987})}\BibitemShut {NoStop}%
\bibitem [{\citenamefont {Agarwal}(1988)}]{Agarwal1988}%
  \BibitemOpen
  \bibfield  {author} {\bibinfo {author} {\bibfnamefont {G.~S.}\ \bibnamefont
  {Agarwal}},\ }\href {\doibase 10.1364/JOSAB.5.001940} {\bibfield  {journal}
  {\bibinfo  {journal} {J. Opt. Soc. Am. B}\ }\textbf {\bibinfo {volume} {5}},\
  \bibinfo {pages} {1940} (\bibinfo {year} {1988})}\BibitemShut {NoStop}%
\bibitem [{\citenamefont {Joshi}\ and\ \citenamefont {Puri}(1990)}]{Joshi1990}%
  \BibitemOpen
  \bibfield  {author} {\bibinfo {author} {\bibfnamefont {A.}~\bibnamefont
  {Joshi}}\ and\ \bibinfo {author} {\bibfnamefont {R.~R.}\ \bibnamefont
  {Puri}},\ }\href {\doibase 10.1103/PhysRevA.42.4336} {\bibfield  {journal}
  {\bibinfo  {journal} {Phys. Rev. A}\ }\textbf {\bibinfo {volume} {42}},\
  \bibinfo {pages} {4336} (\bibinfo {year} {1990})}\BibitemShut {NoStop}%
\bibitem [{\citenamefont {Gerry}\ and\ \citenamefont
  {Welch}(1992)}]{Gerry1992}%
  \BibitemOpen
  \bibfield  {author} {\bibinfo {author} {\bibfnamefont {C.~C.}\ \bibnamefont
  {Gerry}}\ and\ \bibinfo {author} {\bibfnamefont {R.~F.}\ \bibnamefont
  {Welch}},\ }\href {\doibase 10.1364/JOSAB.9.000290} {\bibfield  {journal}
  {\bibinfo  {journal} {J. Opt. Soc. Am. B}\ }\textbf {\bibinfo {volume} {9}},\
  \bibinfo {pages} {290} (\bibinfo {year} {1992})}\BibitemShut {NoStop}%
\bibitem [{\citenamefont {Nahmad-Achar}\ \emph {et~al.}(2015)\citenamefont
  {Nahmad-Achar}, \citenamefont {Cordero}, \citenamefont {Casta\~{n}os},\ and\
  \citenamefont {L\'{o}pez-Pe\~{n}a}}]{NahmadAchar2015}%
  \BibitemOpen
  \bibfield  {author} {\bibinfo {author} {\bibfnamefont {E.}~\bibnamefont
  {Nahmad-Achar}}, \bibinfo {author} {\bibfnamefont {S.}~\bibnamefont
  {Cordero}}, \bibinfo {author} {\bibfnamefont {O.}~\bibnamefont
  {Casta\~{n}os}}, \ and\ \bibinfo {author} {\bibfnamefont {R.}~\bibnamefont
  {L\'{o}pez-Pe\~{n}a}},\ }\href {\doibase 10.1088/0031-8949/90/7/074026}
  {\bibfield  {journal} {\bibinfo  {journal} {Phys. Scripta}\ }\textbf
  {\bibinfo {volume} {90}},\ \bibinfo {pages} {074026} (\bibinfo {year}
  {2015})}\BibitemShut {NoStop}%
\bibitem [{\citenamefont {Cordero}\ \emph {et~al.}(2021)\citenamefont
  {Cordero}, \citenamefont {Nahmad-Achar}, \citenamefont {L\'{o}pez-Pe\~{n}a},\
  and\ \citenamefont {Casta\~{n}os}}]{Cordero2021}%
  \BibitemOpen
  \bibfield  {author} {\bibinfo {author} {\bibfnamefont {S.}~\bibnamefont
  {Cordero}}, \bibinfo {author} {\bibfnamefont {E.}~\bibnamefont
  {Nahmad-Achar}}, \bibinfo {author} {\bibfnamefont {R.}~\bibnamefont
  {L\'{o}pez-Pe\~{n}a}}, \ and\ \bibinfo {author} {\bibfnamefont
  {O.}~\bibnamefont {Casta\~{n}os}},\ }\href {\doibase
  10.1088/1402-4896/abd653} {\bibfield  {journal} {\bibinfo  {journal} {Phys.
  Scripta}\ }\textbf {\bibinfo {volume} {96}},\ \bibinfo {pages} {035104}
  (\bibinfo {year} {2021})}\BibitemShut {NoStop}%
\bibitem [{\citenamefont {Casta\~{n}os}\ \emph {et~al.}(2022)\citenamefont
  {Casta\~{n}os}, \citenamefont {Cordero}, \citenamefont {L\'{o}pez-Pe\~{n}a},\
  and\ \citenamefont {Nahmad-Achar}}]{Castanos2022}%
  \BibitemOpen
  \bibfield  {author} {\bibinfo {author} {\bibfnamefont {O.}~\bibnamefont
  {Casta\~{n}os}}, \bibinfo {author} {\bibfnamefont {S.}~\bibnamefont
  {Cordero}}, \bibinfo {author} {\bibfnamefont {R.}~\bibnamefont
  {L\'{o}pez-Pe\~{n}a}}, \ and\ \bibinfo {author} {\bibfnamefont
  {E.}~\bibnamefont {Nahmad-Achar}},\ }\href {\doibase
  10.1088/1751-8121/aca6bb} {\bibfield  {journal} {\bibinfo  {journal} {J.
  Phys. A: Math. Theor.}\ }\textbf {\bibinfo {volume} {55}},\ \bibinfo {pages}
  {485302} (\bibinfo {year} {2022})}\BibitemShut {NoStop}%
\bibitem [{\citenamefont {L\'{o}pez-Pe\~{n}a}\ \emph
  {et~al.}(2021)\citenamefont {L\'{o}pez-Pe\~{n}a}, \citenamefont {Cordero},
  \citenamefont {Nahmad-Achar},\ and\ \citenamefont
  {Casta\~{n}os}}]{LopezPena2021}%
  \BibitemOpen
  \bibfield  {author} {\bibinfo {author} {\bibfnamefont {R.}~\bibnamefont
  {L\'{o}pez-Pe\~{n}a}}, \bibinfo {author} {\bibfnamefont {S.}~\bibnamefont
  {Cordero}}, \bibinfo {author} {\bibfnamefont {E.}~\bibnamefont
  {Nahmad-Achar}}, \ and\ \bibinfo {author} {\bibfnamefont {O.}~\bibnamefont
  {Casta\~{n}os}},\ }\href {\doibase 10.1088/1402-4896/abd654} {\bibfield
  {journal} {\bibinfo  {journal} {Phys. Scripta}\ }\textbf {\bibinfo {volume}
  {96}},\ \bibinfo {pages} {035103} (\bibinfo {year} {2021})}\BibitemShut
  {NoStop}%
\bibitem [{\citenamefont {Arimondo}\ \emph {et~al.}(1977)\citenamefont
  {Arimondo}, \citenamefont {Inguscio},\ and\ \citenamefont
  {Violino}}]{Arimondo1977}%
  \BibitemOpen
  \bibfield  {author} {\bibinfo {author} {\bibfnamefont {E.}~\bibnamefont
  {Arimondo}}, \bibinfo {author} {\bibfnamefont {M.}~\bibnamefont {Inguscio}},
  \ and\ \bibinfo {author} {\bibfnamefont {P.}~\bibnamefont {Violino}},\ }\href
  {\doibase 10.1103/RevModPhys.49.31} {\bibfield  {journal} {\bibinfo
  {journal} {Rev. Mod. Phys.}\ }\textbf {\bibinfo {volume} {49}},\ \bibinfo
  {pages} {31} (\bibinfo {year} {1977})}\BibitemShut {NoStop}%
\bibitem [{\citenamefont {Blinov}\ \emph {et~al.}(2004)\citenamefont {Blinov},
  \citenamefont {Moehring}, \citenamefont {Duan},\ and\ \citenamefont
  {Monroe}}]{Blinov2004}%
  \BibitemOpen
  \bibfield  {author} {\bibinfo {author} {\bibfnamefont {B.~B.}\ \bibnamefont
  {Blinov}}, \bibinfo {author} {\bibfnamefont {D.~L.}\ \bibnamefont
  {Moehring}}, \bibinfo {author} {\bibfnamefont {L.-M.}\ \bibnamefont {Duan}},
  \ and\ \bibinfo {author} {\bibfnamefont {C.}~\bibnamefont {Monroe}},\ }\href
  {\doibase 10.1038/nature02377} {\bibfield  {journal} {\bibinfo  {journal}
  {Nature}\ }\textbf {\bibinfo {volume} {428}},\ \bibinfo {pages} {153}
  (\bibinfo {year} {2004})}\BibitemShut {NoStop}%
\bibitem [{\citenamefont {Wilk}\ \emph {et~al.}(2007)\citenamefont {Wilk},
  \citenamefont {Webster}, \citenamefont {Kuhn},\ and\ \citenamefont
  {Rempe}}]{Wilk2007}%
  \BibitemOpen
  \bibfield  {author} {\bibinfo {author} {\bibfnamefont {T.}~\bibnamefont
  {Wilk}}, \bibinfo {author} {\bibfnamefont {S.~C.}\ \bibnamefont {Webster}},
  \bibinfo {author} {\bibfnamefont {A.}~\bibnamefont {Kuhn}}, \ and\ \bibinfo
  {author} {\bibfnamefont {G.}~\bibnamefont {Rempe}},\ }\href {\doibase
  10.1126/science.1143835} {\bibfield  {journal} {\bibinfo  {journal}
  {Science}\ }\textbf {\bibinfo {volume} {317}},\ \bibinfo {pages} {488}
  (\bibinfo {year} {2007})}\BibitemShut {NoStop}%
\bibitem [{\citenamefont {Specht}\ \emph {et~al.}(2011)\citenamefont {Specht},
  \citenamefont {Nölleke}, \citenamefont {Reiserer}, \citenamefont {Uphoff},
  \citenamefont {Figueroa}, \citenamefont {Ritter},\ and\ \citenamefont
  {Rempe}}]{Specht2011}%
  \BibitemOpen
  \bibfield  {author} {\bibinfo {author} {\bibfnamefont {H.~P.}\ \bibnamefont
  {Specht}}, \bibinfo {author} {\bibfnamefont {C.}~\bibnamefont {Nölleke}},
  \bibinfo {author} {\bibfnamefont {A.}~\bibnamefont {Reiserer}}, \bibinfo
  {author} {\bibfnamefont {M.}~\bibnamefont {Uphoff}}, \bibinfo {author}
  {\bibfnamefont {E.}~\bibnamefont {Figueroa}}, \bibinfo {author}
  {\bibfnamefont {S.}~\bibnamefont {Ritter}}, \ and\ \bibinfo {author}
  {\bibfnamefont {G.}~\bibnamefont {Rempe}},\ }\href {\doibase
  10.1038/nature09997} {\bibfield  {journal} {\bibinfo  {journal} {Nature}\
  }\textbf {\bibinfo {volume} {473}},\ \bibinfo {pages} {190} (\bibinfo {year}
  {2011})}\BibitemShut {NoStop}%
\bibitem [{\citenamefont {Rosenfeld}(2008)}]{Ediss10030}%
  \BibitemOpen
  \bibfield  {author} {\bibinfo {author} {\bibfnamefont {W.}~\bibnamefont
  {Rosenfeld}},\ }\href {http://nbn-resolving.de/urn:nbn:de:bvb:19-100300}
  {\enquote {\bibinfo {title} {Experiments with an entangled system of a single
  atom and a single photon},}\ } (\bibinfo {year} {2008})\BibitemShut {NoStop}%
\bibitem [{\citenamefont {Sugita}(2003)}]{Sugita2003}%
  \BibitemOpen
  \bibfield  {author} {\bibinfo {author} {\bibfnamefont {A.}~\bibnamefont
  {Sugita}},\ }\href {\doibase 10.1088/0305-4470/36/34/310} {\bibfield
  {journal} {\bibinfo  {journal} {J. Phys. A: Math. Gen.}\ }\textbf {\bibinfo
  {volume} {36}},\ \bibinfo {pages} {9081} (\bibinfo {year}
  {2003})}\BibitemShut {NoStop}%
\bibitem [{\citenamefont {Weiner}(2003)}]{Weiner2003}%
  \BibitemOpen
  \bibfield  {author} {\bibinfo {author} {\bibfnamefont {J.}~\bibnamefont
  {Weiner}},\ }\href@noop {} {\emph {\bibinfo {title} {Cold and Ultracold
  Collisions in Quantum Microscopic and Mesoscopic Systems}}}\ (\bibinfo
  {publisher} {Cambridge University Press},\ \bibinfo {year}
  {2003})\BibitemShut {NoStop}%
\bibitem [{\citenamefont {McAlexander}\ \emph {et~al.}(1996)\citenamefont
  {McAlexander}, \citenamefont {Abraham},\ and\ \citenamefont
  {Hulet}}]{McAlexander1996}%
  \BibitemOpen
  \bibfield  {author} {\bibinfo {author} {\bibfnamefont {W.~I.}\ \bibnamefont
  {McAlexander}}, \bibinfo {author} {\bibfnamefont {E.~R.~I.}\ \bibnamefont
  {Abraham}}, \ and\ \bibinfo {author} {\bibfnamefont {R.~G.}\ \bibnamefont
  {Hulet}},\ }\href {\doibase 10.1103/PhysRevA.54.R5} {\bibfield  {journal}
  {\bibinfo  {journal} {Phys. Rev. A}\ }\textbf {\bibinfo {volume} {54}},\
  \bibinfo {pages} {R5} (\bibinfo {year} {1996})}\BibitemShut {NoStop}%
\bibitem [{\citenamefont {Scherf}\ \emph {et~al.}(1996)\citenamefont {Scherf},
  \citenamefont {Khait}, \citenamefont {Jäger},\ and\ \citenamefont
  {Windholz}}]{Scherf1996}%
  \BibitemOpen
  \bibfield  {author} {\bibinfo {author} {\bibfnamefont {W.}~\bibnamefont
  {Scherf}}, \bibinfo {author} {\bibfnamefont {O.}~\bibnamefont {Khait}},
  \bibinfo {author} {\bibfnamefont {H.}~\bibnamefont {Jäger}}, \ and\ \bibinfo
  {author} {\bibfnamefont {L.}~\bibnamefont {Windholz}},\ }\href {\doibase
  10.1007/BF01437417} {\bibfield  {journal} {\bibinfo  {journal} {Zeitschrift
  für Physik D Atoms, Molecules and Clusters}\ }\textbf {\bibinfo {volume}
  {36}},\ \bibinfo {pages} {31} (\bibinfo {year} {1996})}\BibitemShut {NoStop}%
\bibitem [{\citenamefont {Windholz}\ \emph {et~al.}(1992)\citenamefont
  {Windholz}, \citenamefont {Musso}, \citenamefont {Zerza},\ and\ \citenamefont
  {J\"ager}}]{Windholz1992}%
  \BibitemOpen
  \bibfield  {author} {\bibinfo {author} {\bibfnamefont {L.}~\bibnamefont
  {Windholz}}, \bibinfo {author} {\bibfnamefont {M.}~\bibnamefont {Musso}},
  \bibinfo {author} {\bibfnamefont {G.}~\bibnamefont {Zerza}}, \ and\ \bibinfo
  {author} {\bibfnamefont {H.}~\bibnamefont {J\"ager}},\ }\href {\doibase
  10.1103/PhysRevA.46.5812} {\bibfield  {journal} {\bibinfo  {journal} {Phys.
  Rev. A}\ }\textbf {\bibinfo {volume} {46}},\ \bibinfo {pages} {5812}
  (\bibinfo {year} {1992})}\BibitemShut {NoStop}%
\bibitem [{\citenamefont {Sagle}\ \emph {et~al.}(1996)\citenamefont {Sagle},
  \citenamefont {Namiotka},\ and\ \citenamefont {Huennekens}}]{Sagle1996}%
  \BibitemOpen
  \bibfield  {author} {\bibinfo {author} {\bibfnamefont {J.}~\bibnamefont
  {Sagle}}, \bibinfo {author} {\bibfnamefont {R.~K.}\ \bibnamefont {Namiotka}},
  \ and\ \bibinfo {author} {\bibfnamefont {J.}~\bibnamefont {Huennekens}},\
  }\href {\doibase 10.1088/0953-4075/29/12/023} {\bibfield  {journal} {\bibinfo
   {journal} {J. Phys. B: At. Mol. Opt. Phys.}\ }\textbf {\bibinfo {volume}
  {29}},\ \bibinfo {pages} {2629} (\bibinfo {year} {1996})}\BibitemShut
  {NoStop}%
\bibitem [{\citenamefont {Siddons}\ \emph {et~al.}(2008)\citenamefont
  {Siddons}, \citenamefont {Adams}, \citenamefont {Ge},\ and\ \citenamefont
  {Hughes}}]{Siddons2008}%
  \BibitemOpen
  \bibfield  {author} {\bibinfo {author} {\bibfnamefont {P.}~\bibnamefont
  {Siddons}}, \bibinfo {author} {\bibfnamefont {C.~S.}\ \bibnamefont {Adams}},
  \bibinfo {author} {\bibfnamefont {C.}~\bibnamefont {Ge}}, \ and\ \bibinfo
  {author} {\bibfnamefont {I.~G.}\ \bibnamefont {Hughes}},\ }\href {\doibase
  10.1088/0953-4075/41/15/155004} {\bibfield  {journal} {\bibinfo  {journal}
  {J. Phys. B: At. Mol. Opt. Phys.}\ }\textbf {\bibinfo {volume} {41}},\
  \bibinfo {pages} {155004} (\bibinfo {year} {2008})}\BibitemShut {NoStop}%
\bibitem [{\citenamefont {Araki}\ and\ \citenamefont {Lieb}(1970)}]{Araki1970}%
  \BibitemOpen
  \bibfield  {author} {\bibinfo {author} {\bibfnamefont {H.}~\bibnamefont
  {Araki}}\ and\ \bibinfo {author} {\bibfnamefont {E.~H.}\ \bibnamefont
  {Lieb}},\ }\href {\doibase 10.1007/BF01646092} {\bibfield  {journal}
  {\bibinfo  {journal} {Commun. Math. Phys.}\ }\textbf {\bibinfo {volume}
  {18}},\ \bibinfo {pages} {160} (\bibinfo {year} {1970})}\BibitemShut
  {NoStop}%
\bibitem [{\citenamefont {Horodecki}\ \emph {et~al.}(2001)\citenamefont
  {Horodecki}, \citenamefont {Horodecki},\ and\ \citenamefont
  {Horodecki}}]{Horodecki2001}%
  \BibitemOpen
  \bibfield  {author} {\bibinfo {author} {\bibfnamefont {M.}~\bibnamefont
  {Horodecki}}, \bibinfo {author} {\bibfnamefont {P.}~\bibnamefont
  {Horodecki}}, \ and\ \bibinfo {author} {\bibfnamefont {R.}~\bibnamefont
  {Horodecki}},\ }\enquote {\bibinfo {title} {Mixed-state entanglement and
  quantum communication},}\ in\ \href {\doibase 10.1007/3-540-44678-8_5} {\emph
  {\bibinfo {booktitle} {Quantum Information: An Introduction to Basic
  Theoretical Concepts and Experiments}}}\ (\bibinfo  {publisher} {Springer
  Berlin Heidelberg},\ \bibinfo {address} {Berlin, Heidelberg},\ \bibinfo
  {year} {2001})\ pp.\ \bibinfo {pages} {151--195}\BibitemShut {NoStop}%
\end{thebibliography}%

\end{document}